\documentclass[pdflatex,sn-mathphys-num]{sn-jnl}

\usepackage{graphicx}%
\usepackage{multirow}%
\usepackage{amsmath,amssymb,amsfonts}%
\usepackage{amsthm}%
\usepackage{mathrsfs}%
\usepackage[title]{appendix}%
\usepackage{xcolor}%
\usepackage{textcomp}%
\usepackage{manyfoot}%
\usepackage{booktabs}%
\usepackage{algorithm}%
\usepackage{algorithmicx}%
\usepackage{algpseudocode}%
\usepackage{listings}%

\usepackage[utf8]{inputenc}
\usepackage[normalem]{ulem}

\usepackage{physics}
\usepackage{bm}
\def\bra#1{\langle{#1}|}
\def\ket#1{|{#1}\rangle}

\theoremstyle{thmstyleone}%
\theoremstyle{thmstyletwo}%

\theoremstyle{thmstylethree}%

\begin{document}

\title[Superradiant decay in non-Markovian Waveguide Quantum Electrodynamics]{Superradiant decay in non-Markovian Waveguide Quantum Electrodynamics}

\author[1,2]{\fnm{Rosa Lucia} \sur{Capurso}}

\author[3,4]{\fnm{Giuseppe} \sur{Calaj\'o}}

\author[3,5,4]{\fnm{Simone} \sur{Montangero}}

\author[1,2]{\fnm{Saverio} \sur{Pascazio}}

\author*[1,2]{\fnm{Francesco V.} \sur{Pepe}}\email{francesco.pepe@ba.infn.it}

\author[1,2,6]{\fnm{Maria} \sur{Maffei}}

\author[1,2]{\fnm{Giuseppe} \sur{Magnifico}}

\author[1,2]{\fnm{Paolo} \sur{Facchi}}

\affil[1]{\orgdiv{Dipartimento di Fisica}, \orgname{Universit\`{a} di Bari}, \orgaddress{\city{Bari}, \postcode{I-70126}, \country{Italy}}}

\affil[2]{\orgname{Istituto Nazionale di Fisica Nucleare (INFN) - Sezione di Bari}, \orgaddress{\city{Bari}, \postcode{I-70125}, \country{Italy}}}

\affil[3]{\orgdiv{Dipartimento di Fisica e Astronomia ``G.~Galilei''}, \orgname{Universit\`a di Padova}, \orgaddress{\city{Padova}, \postcode{I-35131}, \country{Italy}}}

\affil[4]{\orgname{Istituto Nazionale di Fisica Nucleare (INFN) - Sezione di Padova}, \orgaddress{\city{Padova}, \postcode{I-35131}, \country{Italy}}}

\affil[5]{\orgdiv{Padua Quantum Technologies Research Center}, \orgname{Universit\`a di Padova}, \orgaddress{\city{Padova}, \postcode{I-35131}, \country{Italy}}}

\affil[6]{\orgname{Universit\'e de Lorraine, CNRS, LPCT}, \orgaddress{\city{Nancy}, \postcode{F-54000}, \country{France}}}


\abstract{An array of initially excited emitters coupled to a one-dimensional waveguide exhibits superradiant decay under the Born–Markov approximation, manifested as a coherent burst of photons in the output field. In this work, we employ tensor-network methods to investigate its non-Markovian dynamics induced by finite time delays in photon exchange among the emitters. We find that the superradiant burst breaks into a structured train of correlated photons, each intensity peak corresponding to a specific photon number.
We quantify the emitter–photon and emitter–emitter entanglement generated during this process and show that the latter emerges in the long-time limit, as part of the excitation becomes trapped within the emitters’ singlet subspace. We finally  consider  the decay of the system’s most radiant state, the symmetric Dicke state, and show that time delay can lead to decay rates exceeding those predicted by the Markovian approximation.}




\maketitle

\section*{Introduction}

The study of the collective emission from ensembles of atoms, coupled to a shared photonic environment, has long been a hallmark of quantum optics and quantum electrodynamics (QED), dating back to Dicke's seminal work~\cite{PhysRev.93.99}. In this work, a fully excited ensemble of two-level quantum emitters, contained within a much smaller distance than their emission wavelength, interacts collectively with the electromagnetic field, giving rise to a synchronized emission process, where the atoms lock in phase as they decay, producing a burst of coherent radiation, known as \textit{superradiance}~\cite{GROSS1982301}. Fast and intense light emission is accompanied by significantly larger decay rates than those of isolated atoms. Superradiant decay has been observed both in dense ensembles of disordered emitters~\cite{PhysRevLett.30.309,PhysRevLett.36.1035,PhysRevLett.39.547,science.285.5427.571,scheibner2007superradiance}, and in cavity QED setups~\cite{mlynek2014observation,PhysRevLett.98.053603,PhysRevLett.49.117,PhysRevLett.131.253603}. Besides affecting the emission rate and intensity of the emitted light, collective decay can also modify correlations in the radiation field~\cite{PRXQuantum.3.010338,jahnke2016giant}, leading to the emergence of second-order coherence~\cite{bach2024emergence,PhysRevLett.134.153602}.

Although the original Dicke scenario focused on the interaction between the electromagnetic field and tightly localized emitters, recent interest has turned to examining how superradiant decay manifests in more general geometries. For sub-wavelength atomic arrays~\cite{PhysRevA.86.031602,PhysRevLett.116.103602,PhysRevA.94.013847,PhysRevX.7.031024,PhysRevLett.126.223602,PhysRevA.104.013701,rui2020subradiant,holzinger2025beyond}, collective effects, enhanced by the ordered arrangement of the emitters, was shown to still support superradiant decay, provided crucial conditions on the dimensionality and lattice geometry of the system are satisfied~\cite{masson2022universality,PhysRevResearch.4.023207,PhysRevResearch.5.013091,PhysRevA.106.053717,holzinger2025collective,mok2024universal,robicheaux2021theoretical,mok2023dicke,malz2022large}.
Among different emitters arrangements, waveguide QED~\cite{shen2007strongly,sheremet2023waveguide,roy2017colloquium} offers an experimentally accessible scenario to study interactions between multiple emitters and light confined within a one-dimensional channel, either at optical~\cite{lodahl2015interfacing,hood2016atom,corzo2019waveguide,prasad2020correlating,tiranov2023collective} or microwave frequencies~\cite{astafiev2010resonance,brehm2021waveguide,mirhosseini2019cavity,kannan2023demand,PhysRevX.12.031036,zhang2023superconducting}. In this setting, superradiant decay has been observed for a few emitters across different experimental platforms~\cite{science.1244324,tiranov2023collective,kim2025super}. For many emitters, the emergence of a superradiant burst has been experimentally observed in a cascade scenario~\cite{PhysRevX.14.011020} and is predicted to induce mirror symmetry breaking with emergent chirality in bi-directional waveguide QED~\cite{PhysRevLett.131.033605}.

All the effects discussed above rely on the Born–Markov approximation~\cite{CohenTannoudji1998}, which neglects time delays arising from the finite speed of photon propagation.
Waveguide QED provides a natural framework to explore atomic decay beyond this approximation, explicitly accounting for such time delays. 
The non-Markovian regime has garnered growing interest in recent years, due to the emergence of novel phenomena and potential applications. The phenomenology associated to the finite propagation times involves generation of photonic cluster states~\cite{pnas.1711003114,ferreira2024deterministic}, excitation of dressed Bound States in the Continuum (BICs)~\cite{PhysRevLett.122.073601,PhysRevA.94.043839,PhysRevA.104.013705,PhysRevA.103.033704,magnifico2025non}, improved single-photon sources~\cite{PhysRevA.110.L031703}, and preparation of genuinely non-Markovian steady states~\cite{PhysRevLett.128.083603}.
To address this complex non-Markovian atom-light dynamics, several approaches were developed, including diagrammatic analytical methods~\cite{PhysRevA.102.013727,PhysRevA.104.023709}, numerical techniques based on tensor networks~\cite{PhysRevLett.116.093601,PhysRevA.93.062104,PhysRevResearch.3.023030,PhysRevX.14.031043,papaefstathiou2024efficient}, and effective model descriptions~\cite{PhysRevX.14.031043,PhysRevA.106.023708,zhang2022embedding,cilluffo2024multimode}. In parallel, significant experimental development, especially in the circuit QED domain, has enabled access to the regimes where non-Markovianity becomes relevant~\cite{andersson2019non, PhysRevX.11.041043, PhysRevLett.131.103603, ferreira2024deterministic}.
Despite recent significant progress, the impact of non-negligible time delays on collective many-body decay remains largely unexplored. After the seminal paper of Milonni and Knight  considering the effect of retardation on a pair of two-level atoms with only one excitation \cite{PhysRevA.10.1096}, most of the efforts were primarily focused on the enhanced superradiant decay involving either few emitters or single-excitation states \cite{PhysRevLett.124.043603,PhysRevResearch.1.032042,PhysRevA.102.043718,PhysRevResearch.6.023213,PhysRevResearch.2.013238}. More recently, non-Markovian many-body decay was investigated in the context of a chiral (i.e., unidirectional) waveguide, where the emergence of an effective maximum number of atoms that can contribute to superradiant dynamics was demonstrated~\cite{PhysRevLett.134.173601}.

In this work, we analyze the effect of an arbitrary photon propagation time on many-body superradiant emission, by characterizing the dynamics of excitation transfer from the emitters to the field and investigating whether non-Markovian effects can shape the intensity and correlations of the emitted photons.
To address this problem, we employ a Tensor Network (TN) method similar to those developed in~\cite{PhysRevLett.116.093601,PhysRevA.93.062104,PhysRevResearch.3.023030,PhysRevX.14.031043}, appropriately extended to a larger number of emitters and propagation times that can be comparable with the excited level lifetime.
Using this tool, we investigate the many-body decay of an array of initially excited emitters perfectly coupled to a one-dimensional waveguide. While the overall superradiant decay rate progressively decreases with increasing propagation time, we observe that for delays shorter than the individual decay time, the emitted superradiant burst breaks up into a train of correlated photons, sequentially ordered by photon number. Such an emission is induced by a \emph{time-delayed phase-locking} process that progressively involves larger portions of the atomic array, as sketched Fig.~\ref{fig:1}.  
The superradiant decay occurs without generating significant entanglement among the emitters, thereby extending recent results obtained in the Markovian limit~\cite{bassler2025absence,rosario2025unraveling,zhang2025unraveling}.
On the other hand, we observe the onset of entanglement among emitters at asymptotic times, much larger than the timescales determined by single-emitter lifetime and photon propagation time. By analyzing the reduced density matrix of the emitter system, we connect this effect with the existence of BICs for our specific parameter choice. Notably, beyond the well-known single-excitation BICs~\cite{PhysRevLett.122.073601,PhysRevA.104.013705,PhysRevA.94.043839,PhysRevA.103.033704,magnifico2025non}, we also observe relaxation towards multi-excitation entangled states.
We finally extend previous results on enhanced superradiant decay in the single-excitation regime~\cite{PhysRevLett.124.043603,PhysRevResearch.1.032042,PhysRevA.102.043718} to the multi-excitation case. Specifically, we study the decay of the system's most radiant state, the symmetric Dicke state with half emitters excited, showing that propagation time leads also in this case to enhanced instantaneous decay rates compared to the Markovian case.

\section*{Results}

\subsection*{Model}\label{Sec.model}

\begin{figure}
    \centering
    \includegraphics[width=0.75\textwidth]{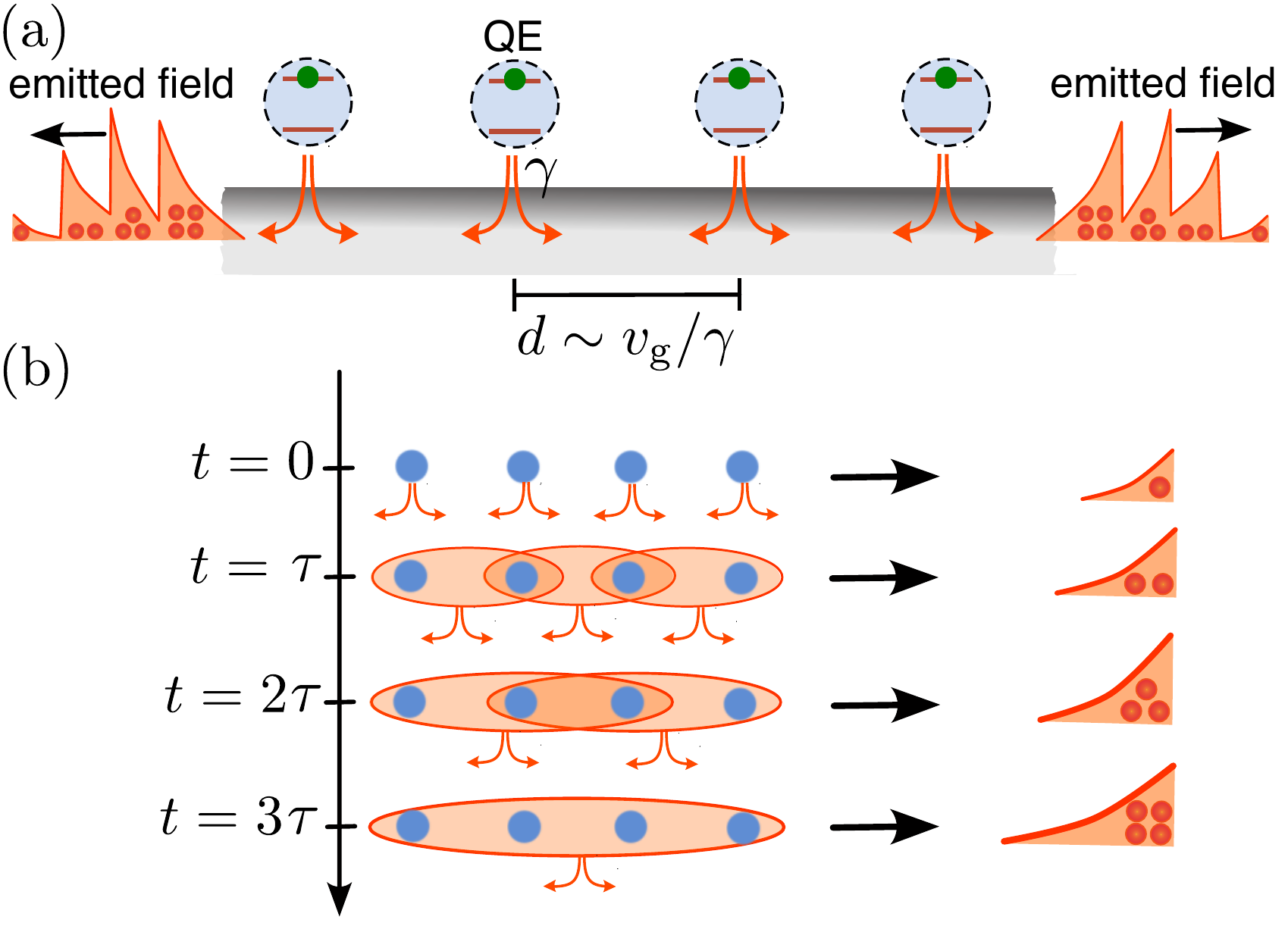}
    \caption{Schematic representation of the main process analyzed in this work. \textit{Panel (a).} Array of $N$ two-level quantum emitters, separated by equal distances $d$, initially prepared in their excited state $\otimes_{j}\ket{e_j}$. The emitted photons can propagate in a waveguide, with the isolated-emitter decay rate $\gamma$ determining the strength of the coupling between each emitter and the field. The propagation time of a photon between two neighboring emitters is $\tau=d/v_g$ with $v_g$ being the photon group velocity. In the Markovian regime, where
    $\eta=\gamma \tau \ll 1/N$, the output field is characterized by a superradiant burst of photons. However, as $\eta$ increases, the emission turns into a train of correlated photon pulses, with each intensity peak corresponding to a specific photon number. \textit{Panel (b)} Graphical representation of the time-delayed phase-locking process, that progressively involves larger portions of the atomic array, at the basis of the phenomenology of superradiance in the non-Markovian regime.} 
    \label{fig:1}
\end{figure}

We consider $N$ quantum emitters of negligible size, coupled to the same linear waveguide at different points, with $d$ their nearest-neighbor spacing, as sketched in Fig.~\ref{fig:1}(a). We assume that the emitters are coupled to the same waveguide transverse mode, neglecting interaction with other modes, either guided or freely propagating, thereby obtaining an effectively one-dimensional field configuration.

The model Hamiltonian
\begin{equation}\label{eq:hamiltonian}
    \hat H=\hat H_0+\hat H_{\rm f}+\hat H_{\rm I}
\end{equation}
consists of three terms describing the free emitters, the free field, and their interactions, respectively.
The emitters are modeled as fixed identical two-level systems with transition frequency $\omega_0$, with their ground and excited states denoted as $\ket{g_{j}}$ and $\ket{e_{j}}$, respectively, with $j=1,\dots,N$. Their free Hamiltonian is thus given by 
\begin{equation}
\hat H_0 = \hbar \sum_{j=1}^N \omega_0\, \hat\sigma_j^\dagger \hat\sigma_j,
\end{equation}
expressed in terms of ladder operators $\hat{\sigma}_j = \ket{g_j} \bra{e_j}$ and their adjoints.
We consider a symmetric field dispersion relation, that  can be safely linearized for right- and left-propagating modes with group velocity $v_{R/L}=\pm v_g$, within the frequency range relevant to the system dynamics.
The free field Hamiltonian is thus
\begin{equation}
    \hat H_{\rm f} =  \hbar \sum_{D = R,L} \int d\omega\, \omega\, \hat a^\dagger_{D}(\omega) \hat a^{\,}_{D}(\omega),
\end{equation}
where the operators $\hat a_{D}(\omega)$ annihilate photons of frequency $\omega$ that propagate towards the right ($D=R$) or the left ($D=L$), and satisfy canonical commutation relations for bosons.

We assume that the emitters are equally coupled to the waveguide, up to a phase factor determined by their different positions, and radiate symmetrically, in such a way that $\gamma$ would be the total decay rate of an isolated emitter. Since we investigate decay dynamics in a weak-coupling regime, we apply a rotating-wave approximation and assume frequency-independent coupling over the relevant bandwidth. It is worth remarking that the non-Markovian effects that we shall analyze throughout the article are unrelated to strong-coupling effects in the interaction between each single emitter and the waveguide. Therefore, the Hamiltonian that describes interactions of the emitter system with the field reads~\cite{PhysRevA.91.042116,PhysRevLett.116.093601,PhysRevX.14.031043}
\begin{equation}\label{eq:Hint}
    \hat H_{\rm I} =\hbar\sum_{D = R,L} \sum_{j=1}^N \sqrt{\frac{\gamma}{4\pi}} \int d\omega\,  e^{-i\omega x_j / v_D}  \hat \sigma_j\, \hat a^\dagger_{D}(\omega)\, + {\rm H.c.},
\end{equation}
where $x_j= (j-1) d$ denotes the position of the $j$-th emitter.

Through their coupling to the waveguide, the emitters can exchange photons. Due to the finite photon speed in the waveguide, the interaction between two nearest-neighbor emitters is generally delayed by the photon \textit{propagation delay}
\begin{equation}
    \tau = \frac{d}{v_g} .
\end{equation}
The dimensionless parameter 
\begin{equation}
    \eta = \gamma \tau ,
\end{equation}
equal to the ratio of the propagation time $\tau$ to the lifetime $\gamma^{-1}$ of an isolated emitter, determines whether the system of emitters can be accurately described within the Born–Markov approximation. Specifically, memory effects due to propagation throughout the whole system can be safely neglected if $\eta \ll 1/N$, leading to the Markovian approximation. However, as soon as $\eta$ increases, non-Markovian effects due to delayed photon exchange become significant and must be explicitly accounted for, as we will discuss in the following.

\subsubsection*{Markovian limit}

When the timescale of the emitter-photon coupling is much longer than the photon propagation time across the entire array ($\eta \ll 1/N$), the light-mediated interactions between emitters can be effectively considered as instantaneous. In this regime, the field degrees of freedom can be adiabatically eliminated by using the Born–Markov approximation. which results in a master equation determining the dynamics of the emitters density operator $\hat \rho(t)$. In the interaction representation, the master equation reads~\cite{Carmichael1999,Breuer2007,sheremet2023waveguide}
\begin{equation}\label{eq:ME_linbl}
  \dot{\hat\rho}(t)=-i[\hat H_S,\hat\rho(t)]+ \sum_{ij}\Gamma_{ij}\left(2 \hat \sigma_j\hat\rho \hat\sigma_i^\dagger -\{\hat\sigma_i^\dagger\hat\sigma_j,\hat\rho\}\right)\,,
  \end{equation}
where 
\begin{equation}\label{eq:Hcoh}
    \hat{H}_{S} = \frac{\gamma}{2} \sum_{i,j} \sin{\left(k_0 |x_i - x_j|\right)} \, \hat{\sigma}_{i}^{\dagger} \hat{\sigma}_{j},
\end{equation}
describes the coherent collective emitter dynamics, with $k_0 = \omega_0 / v_{\rm g}$ being the resonant wavenumber, while the Kossakowski matrix,
\begin{equation}\label{eq_diss_matrix}
    \Gamma_{ij} = \frac{\gamma}{2} \cos(k_0 |x_i - x_j|),
\end{equation}
encodes the collective dissipative effects, with the off-diagonal elements due to the interference between photons propagating from different emitters.
This matrix has only two nonzero (positive) eigenvalues, associated with two collective jump operators
\begin{equation}
    \hat E_{R(L)}=i\sqrt{\frac{\gamma}{2}}\sum_{j} e^{\pm ik_0x_j}\hat\sigma_j,
\end{equation}
which correspond to radiation emission into left- and right-propagating modes~\cite{PhysRevA.88.043806,Caneva_2015}.

The case where the emitter spacing satisfies 
\begin{equation}\label{eq:mirrorconf}
    k_0 d =\omega_0\tau = 2\pi \ell, \quad \text{with } \ell \in \mathbb{N}
\end{equation}
is known as ``mirror configuration''~\cite{Chang_2012} and admits the existence of symmetric BICs in the one-excitation sector~\cite{PhysRevA.100.023834}. If condition~\eqref{eq:mirrorconf} holds, Eq.~\eqref{eq:ME_linbl} corresponds to the original Dicke model~\cite{PhysRev.93.99}
\begin{equation}\label{eq:ME_Dicke}
  \dot{\hat\rho}(t)= 2 \gamma \left(2 \hat S\hat\rho \hat S^\dagger -\{\hat S^\dagger\hat S,\hat\rho\}\right)\,,
\end{equation}
where $\hat S=\sum_i\hat \sigma_i/2$ is the collective spin operator. In this configuration, a system of even $N$ two-level emitters, that is $N$ spins $1/2$, reduces to a single collective spin $S$, with $S=0,1,\dots,N/2$. The evolution determined by~\eqref{eq:ME_Dicke} does not mix different total spin manifolds
\begin{equation}\label{eq:spinmanif}
    \{ \ket{S,S_z} \}_{-S \leq S_z \leq S} ,
\end{equation}
where the spin projection along the quantization axis is related to the number of excitations $N_{\mathrm{exc}}$ inside the emitter system by $S_z = N_{\mathrm{exc}}-N/2$.

\subsubsection*{Non-Markovian effects}\label{Appendix:B}

Beyond the Markovian limit, the interaction Hamiltonian \eqref{eq:Hint} determines effects that generally mix different spin multiplets. For its relevance to the present work, let us consider the case of the initial state $\ket{\Psi(0)}=\ket{S=N/2,S_z=N/2}\otimes\ket{\mathrm{vac}}$, belonging to the highest-spin multiplet, on which the interaction Hamiltonian acts as
\begin{equation}\label{eq:HPsi0_app}
    H_{\mathrm{I}} \ket{\Psi(0)} = \hbar\sqrt{\frac{\gamma}{4\pi}} \sum_{j=1}^N \ket{g_j^{(1)}} \otimes \int d\omega \, \xi(\omega) \Bigl( e^{-i\omega\tau(j-1)}\ket{\omega_R} + e^{i\omega\tau(j-1)}\ket{\omega_L}  \Bigr) ,
\end{equation}
where $\ket{\omega_D}={\hat{a}}^{\dagger}_D \ket{\mathrm{vac}}$ are single-photon states, 
\begin{equation}
    \ket{g_j^{(1)}} = \bigotimes_{p=1}^{j-1} |e_p\rangle \otimes \ket{g_j} \otimes \bigotimes_{q=j+1}^N |e_q\rangle
\end{equation}
are elements of the factorized $(N-1)$-excitation basis, and $\xi(\omega)$ is a form factor with $\xi(\omega_0)=1$, introduced to regularize integrals.

In the case $\tau=0$, the right-hand side of Eq.~\eqref{eq:HPsi0_app} factors into a single-photon state and the fully symmetric combination of $(N-1)$-excited states
\begin{equation}
    \ket{\Psi_s^{(1)}} = \frac{1}{\sqrt{N}} \sum_{j=1}^N \ket{g_j^{(1)}} ,
\end{equation}
belonging to the $S=N/2$ manifold. To check if, for a finite value of $\tau$, $H_{\mathrm{I}}$ can drive $\ket{\Psi(0)}$ out of the $S=N/2$ multiplet, namely to $(N-1)$-excitation states orthogonal to $\ket{\Psi_s^{(1)}}$, let us consider the projector
\begin{equation}
    P_{s(1)}^{\perp} = \left( I_0 - \ket{\Psi_s^{(1)}} \bra{\Psi_s^{(1)}} \right) \otimes I_{\mathrm{f}} ,
\end{equation}
with $I_0$ and $I_{\mathrm{f}}$ the identities on the emitter and field Hilbert spaces, respectively. The action of $P_{s(1)}^{\perp}$ on $H_{\mathrm{I}} \ket{\Psi(0)}$ yields the vector
\begin{multline}
    P_{s(1)}^{\perp} H_{\mathrm{I}} \ket{\Psi(0)} \\ = \hbar\sqrt{\frac{\gamma}{4\pi}} \sum_{j=1}^N \ket{g_j^{(1)}} 
    \otimes \int d\omega \, \xi(\omega) \Bigl( e^{-i\omega\tau(j-1)} F_{j}(\omega\tau) \ket{\omega_R} + e^{i\omega\tau(j-1)} F_{j}^* (\omega\tau) \ket{\omega_L}  \Bigr) ,
\end{multline}
with $F_{j}(\omega\tau) = 1 - N^{-1} \sum_{k=1}^N e^{i\omega\tau(j-k)} $.
Its squared norm thus reads
\begin{equation}
    \left\lVert P_{s(1)}^{\perp} H_{\mathrm{I}} \ket{\Psi(0)} \right\rVert^2 = \frac{\hbar\gamma}{2\pi} \int d\omega \, \xi^2(\omega) \mathcal{F}(\omega\tau) ,
\end{equation}
where $\mathcal{F}(\omega\tau)=\sum_j |F_{j}(\omega\tau)|^2$ is a positive function for all $N\geq 2$, vanishing only at the points 
\begin{equation}
    \omega\tau = 2\pi \ell', \quad \text{with } \ell' \in \mathbb{Z} 
\end{equation}
and scaling like $\tau^2$ for small delays. Thus, while the asymptotic (on-shell) transition rate towards states outside the fully symmetric manifold vanishes in the mirror configuration \eqref{eq:mirrorconf}, being proportional to
\begin{equation}
    \int d\omega \, \xi^2(\omega) \mathcal{F}(\omega\tau) \delta (\omega-\omega_0) = \mathcal{F}(2\pi\ell) = 0 ,
\end{equation}
the transition amplitude towards these states never vanishes, generally determining non-Markovian effects.

\subsubsection*{Time-delayed photon exchange}

To treat the non-Markovian effects deriving from a finite propagation time, it is convenient to rewrite the interaction Hamiltonian from Eq.~\eqref{eq:Hint} in the interaction picture with respect to the bare emitter and field Hamiltonian, $H_0+H_{\rm f}$, yielding
\begin{align}\label{eq:V_continuousTime}
    \hat{H}^{\mathrm{int}}_{I}(t)= \hbar \sqrt{\frac{\gamma}{2}}\sum_j 
     \hat\sigma_{j} \left[e^{-i \phi_j} \hat b^{\dagger}_{R}\left(t-\tau_j\right)+ e^{i \phi_j} \hat b^{\dagger}_{L}\left(t+\tau_j\right)\right] + \mathrm{H.c.} ,
\end{align}
where $\tau_j= x_j/v_g = (j-1)\tau$, 
\begin{equation}
    \phi_j =\omega_0 \tau_{j}=k_0d(j-1)
\end{equation} 
is the phase associated with photon emission and absorption by the $j$-th emitter, and 
\begin{equation} \label{eq:noise_operators}
   \hat b_{D}(t)= \frac{1}{\sqrt{2\pi}}\int d\omega\, \hat a_{D}(\omega)e^{-i(\omega-\omega_0)t}  
\end{equation}
are the quantum noise operators~\cite{gardiner2004quantum}, satisfying the canonical commutation relations $[\hat b_{D}(t),\hat b^{\dagger}_{D'}(t')]=\delta_{D,D'}\delta(t-t')$ with $D,D'\in\lbrace R,L\rbrace$.

The dynamics described by Eq.~\eqref{eq:V_continuousTime} is notoriously challenging to address in the regime of non-negligible propagation delay $\tau$. Indeed, at each time, the coupling between the atomic array and the field involves multiple quantum noise operators, thus forbidding the standard Born-Markov derivation of the master equation~\eqref{eq:ME_linbl}. To numerically tackle the dynamical problem, it is convenient to coarse-grain the evolution in the spirit of the collision model approach~\cite{gardiner2004quantum,Ciccarello_2017,CICCARELLO20221,Maffei2022,Maffei2024}.
By introducing a coarse-grained time interval $\Delta t \ll \gamma^{-1}$, with $\tau/\Delta t$ a large integer, we define the quantum noise increment operators as $\hat B_{n,D} = (\Delta t)^{-1/2} \int_{n \Delta t}^{(n+1)\Delta t} \mathrm{d}t'\, \hat b_{D}(t')$, where $n$ is an integer. These operators annihilate an excitation in the $n$-th discrete-time field mode, or \textit{collision unit}, propagating in the direction $D$.
 Hence, at the first order $\gamma\Delta t$, the coarse-grained evolution of the system is given by
\begin{equation}\label{eq:collision_Un}
    \ket{\Psi(t_{n+1})} = \exp\left( -i  \sum_j \hat O_{j}(t_n)  \right) \ket{\Psi(t_{n})} ,
\end{equation}
with
\begin{align} \label{eq:longrange}
    \hat O_{j}(t_n)&=\sqrt{\frac{\gamma \Delta t}{2}} \hat\sigma_{j} \left[e^{-i \phi_j} \hat B^{\dagger}_{n-s_j,R}+ e^{i \phi_j} \hat B^{\dagger}_{n+s_j, L}\right]+ \mathrm{H.c.} ,
\end{align}
where we introduced the integers $s_j=(j-1)\tau/\Delta t$. The map~\eqref{eq:collision_Un} evolves the state by coupling the emitter to different collision units at each discrete time step. Its structure inherently captures memory effects arising from the finite propagation delay between emitters: the $n$-th right-propagating collision unit, which interacts with the first emitter at time $t = t_n$, will interact with the $j$-th emitter at time $t = t_n + \tau_j = t_{n + s_j}$. Similarly, the $n$-th left-propagating collision unit that interacts with the first emitter at $t = t_n$ has already interacted with the $j$-th emitter at $t = t_n - \tau_j = t_{n - s_j}$.

\subsubsection*{MPS simulation}\label{Sec_MPS}

The time discretization discussed in the previous section makes the non-Markovian dynamics suitable to take on with a tensor network approach, following and extending the methodology introduced in~\cite{PhysRevLett.116.093601}. In particular, the full state of the joint emitter-field system can be encoded into a collision Matrix Product State (MPS) representation, where specific MPS sites host the emitters, whereas all other sites encode the photonic collision units \cite{PhysRevA.104.L031701,lp1b-yswm}. In this framework, we consider an \textit{even} number $N$ of emitters and group them into $N/2$ sites, each site hosting a pair of emitters that interact with the same collision units at each time step. The photonic sites encode both left- and right-propagating modes and are arranged in the MPS to represent the photon propagation delay. The algorithm drives the evolution of the joint emitter-field system state once the considered initial state is initialized in MPS form. This evolution is carried out by iteratively applying to the MPS the quantum gates that constitute the unitary map in Eq.~\eqref{eq:collision_Un}.

For our simulations, we employ MPS bond dimensions chosen to capture, in the different scenarios considered, the scaling of entanglement with the number of emitters $N$ and the delay parameter $\eta$. Technical details of the MPS implementation, the evolution algorithm, the initial state preparation and the convergence analysis are provided in ``\nameref{Appendix:A}''.

\begin{figure} 
    \centering
    \includegraphics[width=\textwidth]{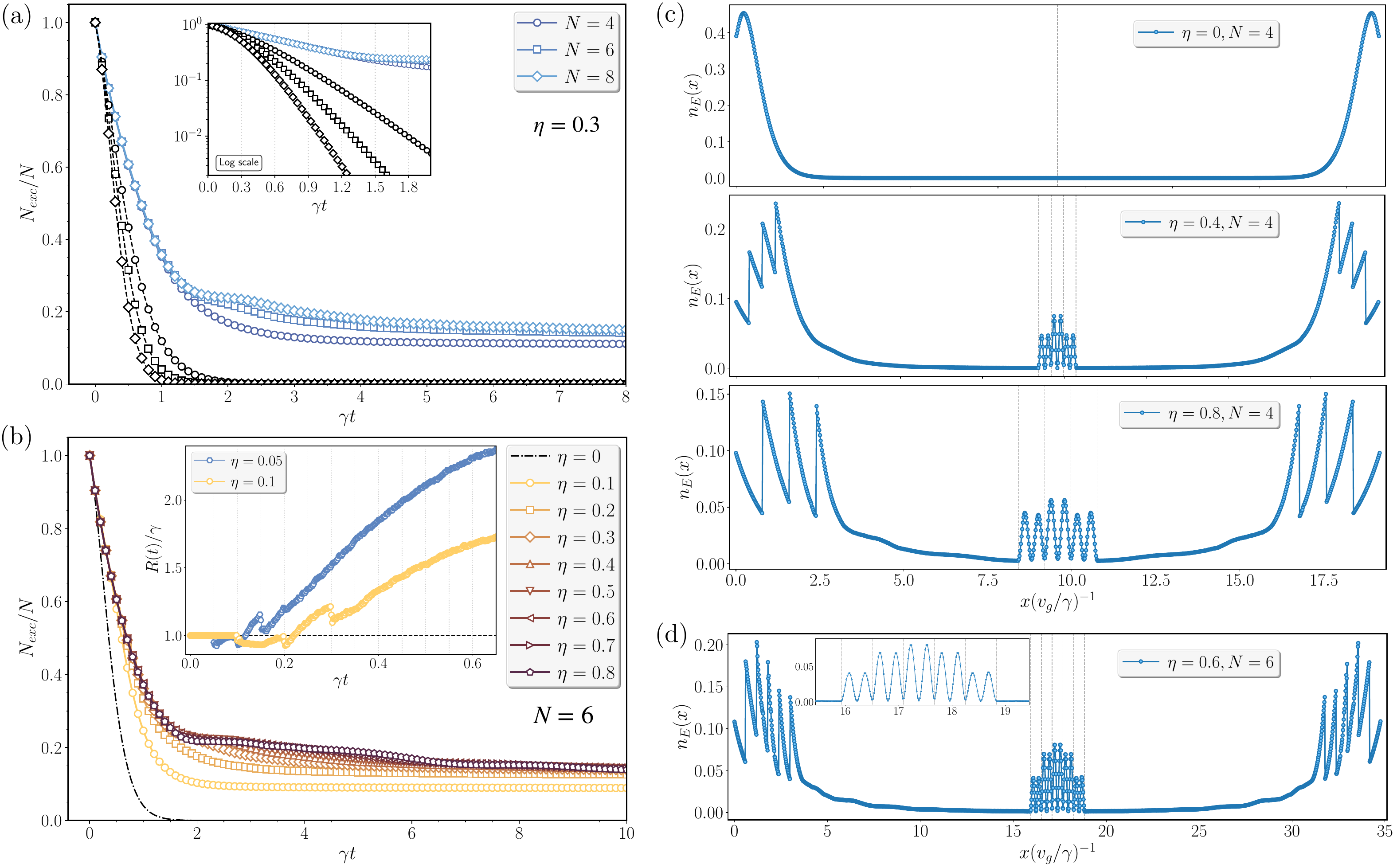}
    \caption{Evolution of emitter excitation number and field energy. \textit{Panels (a)--(b).} Time evolution of the emitter excitation number $N_{\mathrm{exc}}$, defined  in Eq.~\eqref{eq_pmean} and normalized to the number $N$ of emitters, from the $N$-excitation initial state~\eqref{eq:Nexc}. The quantity is plotted in panel (a) for different numbers of emitters with fixed $\eta=0.3$, and in panel (b) for different values of $\eta$ with fixed $N=6$, where the non-Markovianity parameter $\eta=\gamma\tau$ represents the ratio of the propagation delay to the single-emitter lifetime. In both panels, the black dashed lines represent for comparison the behavior in the Markovian limit, which is obtained by solving the emitters' master equation~\eqref{eq:ME_Dicke}. The inset in panel (a) displays the corresponding dynamics on a logarithmic scale. The inset in panel (b) shows the instantaneous decay rate $R(t)$, defined in Eq.~\eqref{eq:rate}, in units of the isolated emitter's decay rate $\gamma$. The dashed horizontal line $R(t)/\gamma=1$ represents the expectation for independent emitters. \textit{Panels (c)--(d).} Spatial profile of the field energy density along the waveguide in the long-time limit $t\approx 9\gamma^{-1}$ for (c) $N=4$ with $\eta=0,0.4,0.8$,  and at $t=16\gamma^{-1}$ for (d) $N=6$ and $\eta=0.6$. 
    The vertical dashed lines indicate the positions of the emitters. The inset in panel (d) provides a magnified view of the energy profile of the field trapped between the emitters.}
    \label{fig:figure_2}
\end{figure}

\subsection*{Non-Markovian superradiant decay}\label{Sec.noMark}

In this section, we investigate the superradiant decay of an array of initially excited emitters coupled to a one-dimensional waveguide in the presence of a non-negligible propagation delay. 
We will focus on frequencies and distances that satisfy the mirror configuration condition~\eqref{eq:mirrorconf}, in order to enhance the deviations from the original Dicke scenario described by Eq.~\eqref{eq:ME_Dicke}. 
This choice is motivated by the possible presence of BICs in the spectrum, which, on the one hand, can lead to significant long-time deviations in the evolution~\cite{PhysRevA.94.043839}, and, on the other hand, can reduce the number of available decay channels, thereby enhancing the decay rate~\cite{Maffei2024}. As a consequence, at least for short time scales, the dynamics is dominated by dissipation, while coherent photon-mediated interactions affect long-time behavior~\cite{PhysRevLett.131.033605}. 
In the following, we consider an even number $N\ge 4$ of emitters, showing how non-negligible propagation delays can provide significant deviations from the Markovian predictions and give rise to unexpected features in the collective emission.

\subsubsection*{Collective decay}

To study collective decay in the presence of propagation delay, we consider an initial state in which all atoms are excited and the waveguide is in the vacuum state $|{\rm vac}\rangle$, namely 
\begin{equation}\label{eq:Nexc}
    \ket{\Psi(0)}=\bigotimes_{j=1}^N |e_j\rangle \otimes|{\rm vac}\rangle .
\end{equation}
In general, the evolution entailed by the model Hamiltonian~\eqref{eq:hamiltonian} preserves symmetry for reflections around the midpoint of the emitter system. This condition, in the Markovian regime, would constrain the dynamics in the manifold $\{ \ket{S=N/2,S_z} \}_{-N/2\leq S_z \leq N/2}$ of the totally symmetric combination of $N$ spins $1/2$, where $S_z=N/2$ coincides with the state of the emitters in~\eqref{eq:Nexc} and $S_z=-N/2$ represents the global ground state. The Markovian evolution results in an enhanced decay rate $\gamma N$  and in a superradiant burst in the intensity of the emitted field, which scales like $N^2$~\cite{PhysRevLett.131.033605}.

The state \eqref{eq:Nexc} is let evolve as $\ket{\Psi(t)} = e^{-i\hat{H}t/\hbar} \ket{\Psi(0)}$ using the coarse-grained dynamics defined by Eq.~\eqref{eq:collision_Un}, employing the MPS methods described in ``\nameref{Sec_MPS}''. 
We consider finite propagation delays characterized by the dimensionless parameter $\eta = \gamma \tau$, while keeping the accumulated phase fixed at integer multiples of $2\pi$, namely $\phi_j = 2\pi (j-1)$, with $j$ being the index that labels the emitter positions. While the Markovian dynamics yielded by Eq.~\eqref{eq:ME_Dicke} is limited to the $S=N/2$ multiplet, manifolds with $S<N/2$ are generally accessible to the exact Hamiltonian dynamics, for any positive value of $\tau$. As shown in ``\nameref{Appendix:B}'', this is related to the fact that, despite the vanishing golden-rule decay rates, the interaction Hamiltonian can still drive the initial state outside the $S=N/2$ multiplet. Such a property is at the basis of the BIC generation protocol defined in Ref.~\cite{PhysRevResearch.6.023213} for $N=2$.

The first quantity we employ to characterize the properties of $\ket{\Psi(t)}$ is the emitter excitation number
\begin{equation}\label{eq_pmean}
    N_{\mathrm{exc}} (t)= \bra{\Psi(t)} \sum_{j=1}^N \hat\sigma_j^\dagger\hat\sigma_j \ket{\Psi(t)} .
\end{equation}
The results are illustrated in Fig.~\ref{fig:figure_2}(a) and (b). In panel (a) we show the evolution of $N_{\mathrm{exc}}(t)$ obtained by fixing $\eta = 0.3$ and varying the system size up to $N = 8$ emitters, while in panel (b) we fix $N = 6$ and vary the parameter $\eta$.
For comparison, we also include the corresponding Markovian limit $\eta\to 0$, computed using the master equation~\eqref{eq:ME_linbl}. The first noticeable effect of non-Markovianity is a reduction in the overall decay rate. However, as we show below, longer propagation delays lead to more profound changes in the collective decay dynamics.

In the Markovian regime, the photon-mediated interaction between the emitters is practically instantaneous, allowing for phase synchronization and collective decay, with a rate that scales as the system size. In contrast, when the interaction is delayed, decay occurs through a process of progressive synchronization. To further investigate this aspect, we analyze the instantaneous decay rate
\begin{equation}\label{eq:rate}
    R(t) = -\frac{1}{N_{\mathrm{exc}}(t)} \frac{d N_{\mathrm{exc}}(t)}{dt} ,
\end{equation}
and plot it for the first delayed curves in the inset of Fig.~\ref{fig:figure_2}(b), where the spontaneous emission rate $\gamma$ (dashed line) is reported for comparison. 
For times \( t < \tau \), the emitters do not interact and decay independently at a rate \( \gamma \). At \( t = \tau \) (i.e.\ $\gamma t = \eta$), the first emitted photons reach the nearest-neighbor emitters, inducing stimulated emission in addition to spontaneous emission. At the small values of $\eta$ reported in the inset of Fig.~\ref{fig:figure_2}(b), interference effect between the two processes leads to a slight decrease in the instantaneous decay rate.
At \( t = 2\tau \) (i.e.\ $\gamma t = 2\eta$), after a sudden slight decrease, $R(t)$ begins to increase, exceeding the spontaneous emission rate $\gamma$. 
This behavior is driven by the backscattering establishing in-phase collective emission between nearest-neighbor emitter pairs, which also includes a superradiant component.
As time evolves, this gradual build-up of collective decay extends beyond nearest-neighbor pairs, progressively involving emitter triplets, and eventually the entire array, as schematically illustrated in Fig.~\ref{fig:1}(b).
For a fixed number of emitters, decreasing the delay leads to a steeper increase of the instantaneous decay rate, ultimately approaching the superradiant limit $N\gamma$ as $\eta \to 0$.

This time-delayed phase-locking process results in two main effects that distinguish this scenario from the Markovian case. 
First, while for $\eta\to 0$ the initial emitter excitations are entirely transferred to the field that freely propagates in the waveguide, for finite $\eta$ part of the excitation remains trapped inside the emitter system, shared by the emitters and a stationary field, as shown in the central parts of Fig.~\ref{fig:figure_2}(c)--(d). This effect is due to the presence of BICs in the spectrum of the system. In the Markovian case, these states are characterized by pure emitter excitation, and since they do not belong to the totally symmetric spin manifold, they are inaccessible to the state evolution. Instead, for finite $\eta$, BICs become dressed with field excitation~\cite{magnifico2025non}, which allows time-delayed multi-photon scattering processes to populate them~\cite{PhysRevLett.122.073601}. This effect is evidenced by the saturation of the average excitation number in the long-time limit and by pronounced oscillations that occur for large delays and extended arrays. These oscillations reflect the slow relaxation into the bound states via multiple scattering events within the array. We shall provide further details on this process in ``\nameref{SecBIC}''.

\subsubsection*{Field energy}

A second consequence of the propagation delay is the alteration of the intensity and correlations of the output field. To characterize the emitted field, we exploit the linearization of the dispersion relation and define the canonical field operators in the space domain as
\begin{equation}
    \hat a_{L}(x)=\frac{1}{\sqrt{v_g}}e^{ik_0 x}
    \hat b_{L}\left(\frac{x}{v_g} \right), \quad \hat a_{R}(x)=\frac{1}{\sqrt{v_g}}e^{-ik_0 x}
    \hat b_{R}\left(-\frac{x}{v_g} \right).
\end{equation}
We thus compute the energy density of the waveguide field as
\begin{equation}
    n_E(x,t)=\sum_{D=R,L} \bra{\Psi(t)}\hat a^\dagger_D(x)\hat a_D(x)\ket{\Psi(t)},
\end{equation}
and we plot in Fig.~\ref{fig:figure_2}(c) its spatial profile in the long-time limit for $N=4$ and different time delays.

In the Markovian regime (upper panel), a single superradiant burst is clearly visible.
The presence of non-negligible propagation delays gives rise to two distinct effects. First, as discussed in the previous section, part of the excitation remains trapped in BICs.
This manifests itself as a portion of the field energy that remains confined between the first and the last emitters, in a standing-wave profile, as shown more clearly in the inset of Fig.~\ref{fig:figure_2}(d). Interestingly, the trapped photons do not correspond solely to single-excitation BICs, but also include contributions from more complex entangled states residing in higher-excitation subspaces, as we will illustrate in ``\nameref{SecBIC}''.
The second observable effect is the progressive splitting of the original superradiant burst into multiple peaks, whose separation increases as the delay grows.

While, in the Markovian regime, all the atoms contribute together to the output intensity, even at its wavefronts, for finite $\eta$ the output field consists until $t < \tau$ solely of an exponentially shaped wave packet emitted by the atom in the chain that is closer to the boundary. At later times, the progressive synchronization process discussed in the previous section comes into play, manifesting as a global maximum in the intensity profile reminiscent of the Markovian superradiant burst.
For finite but small $\eta$, this maximum coincides with the final peak, associated with the collective decay of the entire ensemble, as shown in the second panel in Fig.~\ref{fig:figure_2}(c). As the time delay increases, the maximum shifts to earlier peaks, corresponding to emission from sub-portions of the chain, as in the third panel of Fig.~\ref{fig:figure_2}(c)]. This highlights a transition towards the limit in which the field is radiated by essentially independent emitters.

\begin{figure}
    \centering
    \includegraphics[width=\textwidth]{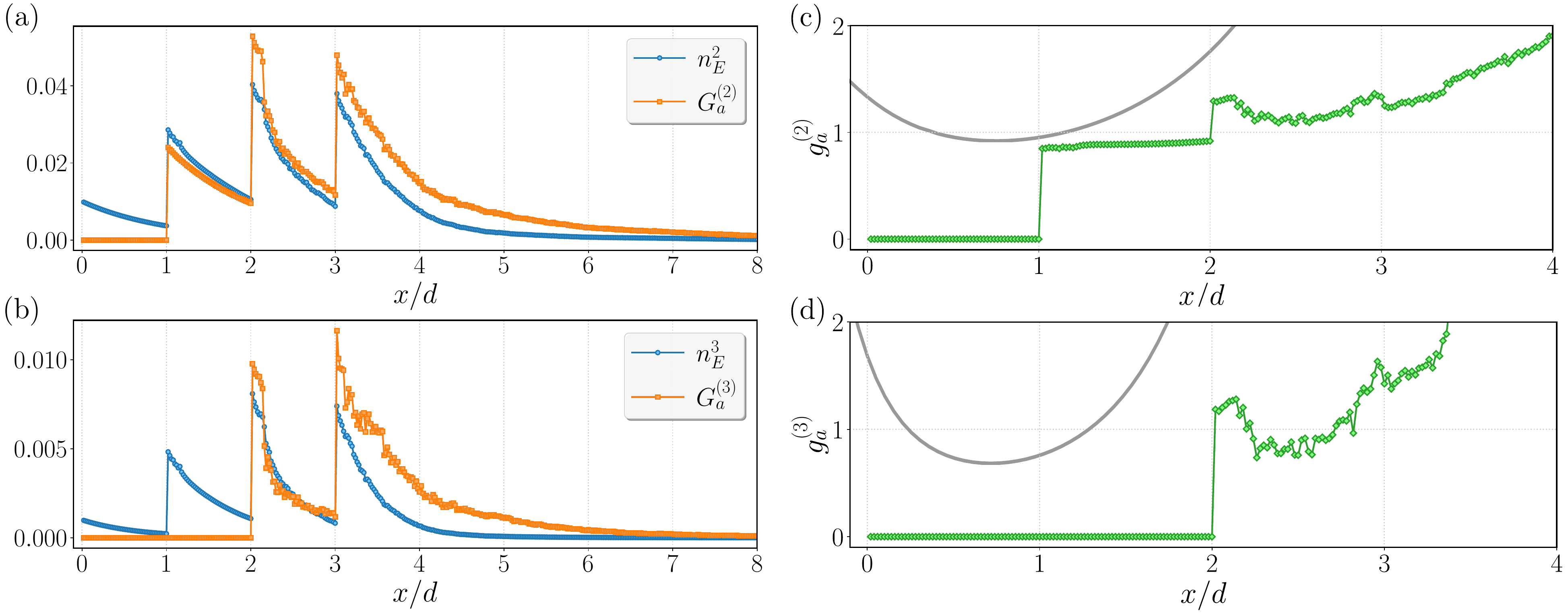}
    \caption{Auto-correlation profiles of the emitted field. \textit{Panels (a)-(b).} Spatial profiles of the second- and third-order auto-correlation functions, \( G_a^{(2)} \) and \( G_a^{(3)} \), as defined in Eq.~\eqref{eq_correlation}, together with the corresponding \( m \)-th power of the intensity, are shown for \( N = 4 \) emitters and time delay \( \eta = 0.5 \). All profiles correspond to a snapshot at time $t_0=8\tau$ of the left part of the waveguide between the left edge and the first emitter. \textit{Panels (c)-(d).} Spatial profiles of the corresponding normalized  auto-correlation functions \( g_a^{(m)} \). In gray we show the correlations computed at zero time delay, $\eta=0$, corresponding to the  Markovian regime.}\label{fig:figure_3}
\end{figure}

\begin{figure} 
    \centering
    \includegraphics[width=\textwidth]{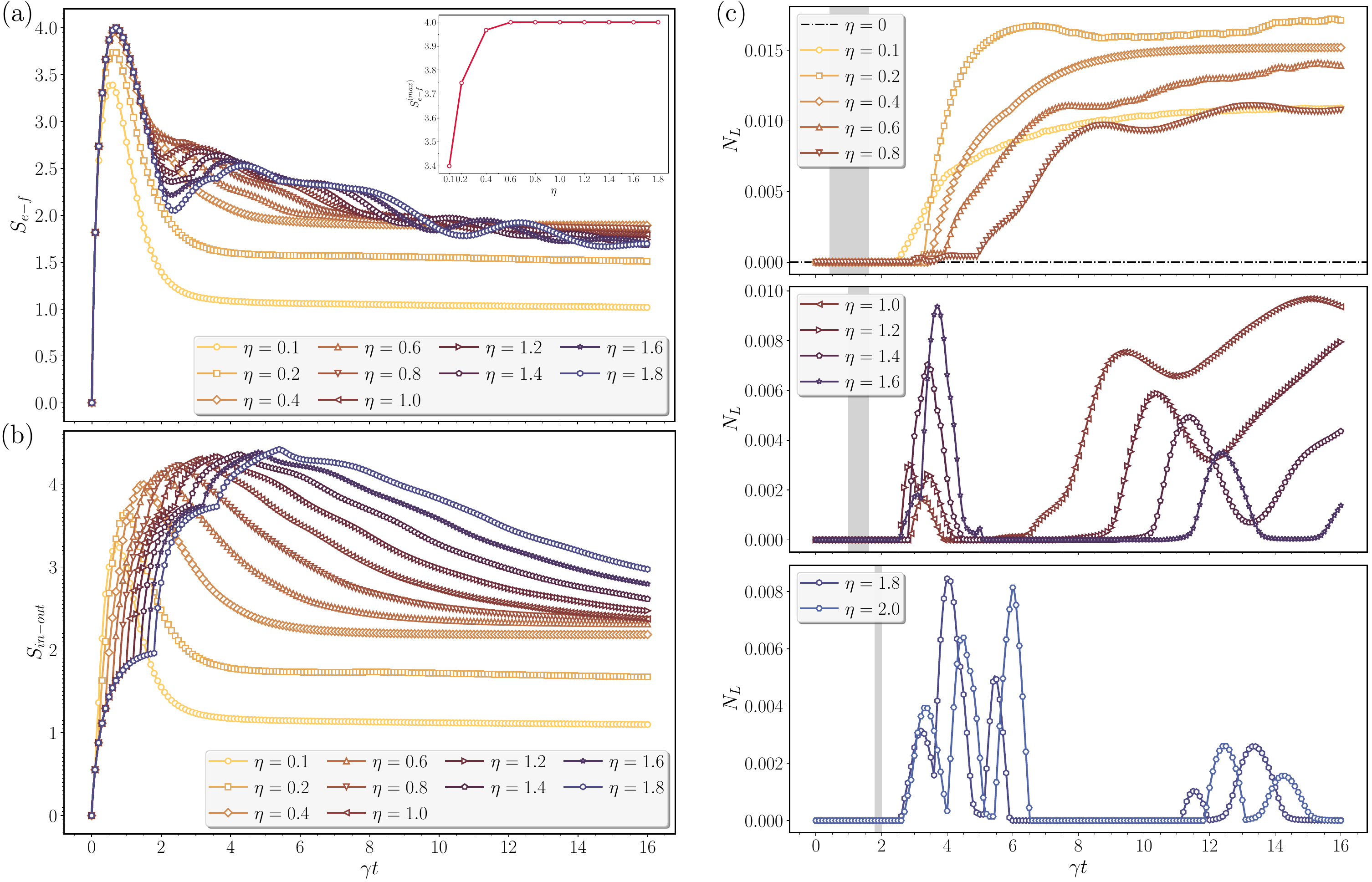}
    \caption{Time evolution of the entanglement generated from the initial state~\eqref{eq:Nexc} for $N = 4$ and different values of $\eta=\gamma\tau$. Panel (a) shows the atom–field entanglement entropy $S_{\mathrm{e-f}}$; panel (b) the entanglement entropy $S_{\mathrm{in-out}}$ between the trapping region and the outer parts of the waveguide; panel (c) the logarithmic negativity $N_L$ computed for half of the emitter chain. The gray box indicates the time window during which the maximum of the output intensity occurs for the corresponding values of $\eta$.}\label{fig:figure_4}
\end{figure}

\subsubsection*{Photon correlations} \label{sec:photon_corr}

To get more insight into this time-delayed collective emission we compute the photon correlations produced in this process.
Photon correlations encompass two types of contributions: auto-correlations, which occur between photons exiting the same channel (either left or right), and cross-correlations, observed between photons emitted in the two  channels. In the following, we will focus on the analysis of the auto-correlation functions due to their experimental accessibility and easier extension to higher-order. It is worth noting, however, that in the limit of the Dicke model (as described in Eq.~\eqref{eq:ME_Dicke}), the auto and cross-correlation contributions become equivalent.
We thus compute the second- and third-order photon auto-correlation functions for photon emitted in the same direction as
\begin{equation}\label{eq_correlation}
G_a^{(m)}(x,t)=\sum_{D=R,L} \bra{\Psi(t)}[\hat a^\dagger_D(x)]^m[\hat a_D(x)]^m\ket{\Psi(t)},
\end{equation}
where the index $m$ indicates the correlation order.
In Fig.~\ref{fig:figure_3}(a) and (b) we plot the profile of these functions on the left edge of the waveguide  for $N=4$ and  $\eta=0.5$ along with the corresponding $m$-th power of the intensity $n_E^m$. The normalized auto-correlation function, $g_a^{(m)}=G_a^{(m)}/n_E^{m}$, reveals the presence of bunching or anti-bunching in the emitted field as it would be measured by a detector placed on the left edge of the waveguide.
For $g_a^{(m)}>1(g_a^{(m)}<1)$ the  emitted photons are bunched (anti-bunched), when $g_a^{(m)}=1$ the photons instead are  uncorrelated. The profile of the normalized correlation function of the left portion of the waveguide, evaluated at the same time $t_0 = 8 \tau$ as in Fig.~\ref{fig:figure_3}~(a)--(b), is shown in Fig.~\ref{fig:figure_3}~(c)--(d). This is the photon propagation time from the first emitter and the detector situated at the left edge of the waveguide.

Figure~\ref{fig:figure_3} shows that each intensity peak corresponds to the emission of up to a specific number of photons, determined by the number of emitters located within the same light cone (see schematic in Fig.~\ref{fig:1}). During the decay, photon correlations are progressively established through the interplay between stimulated emission and backscattering, ultimately becoming associated with the observed intensity peaks.
 We now examine the photon correlations in each peak in detail.

\begin{itemize}

 \item  Between $x=0$ and $x=d$, we observe the field that reaches the detector between times $t_0$ and $t_0+\tau$. This portion of the field has $G_a^{(m)}= g_a^{(m)}=0$, since it contains only a single-photon component, corresponding to the emission from the closest atom to the boundary.

    \item  Between $x=d$ and $x=2d$,  we observe the field that reaches the detector within the time interval $t_0+\tau$ to $t_0+2\tau$.  
    Here the emitted field originates from only two emitters entering the same light cone. As a result, it does not contain any three-photon component and exhibits only a finite two-photon contribution, $G_a^{(2)} \neq 0$.
    This contribution arises from the stimulated emission of the first atom, which has not yet fully decayed, induced by the photon emitted from the second atom~\cite{PhysRevLett.108.143602}.  This process generates a pair of uncorrelated photons, as also occurs for two emitters in the Markovian limit~\cite{PhysRevLett.131.033605}, as indicated by the value $g_a^{(2)} \approx 1$. 
     The slight reduction below one is due to the probability that the photon emitted by the second atom encounters the first atom in its ground state and is reflected, thereby reducing the probability of detecting two photons simultaneously and consequently lowering the value of~$g_a^{(2)}$.

    \item Between $x = 2d$ and $x = 3d$, we observe the field that reaches the detector between times $t_0+2\tau$ and $t_0+3\tau$, which contains two- and three-photon components, namely $G_a^{(2)} \neq 0$ and $G_a^{(3)} \neq 0$. In this portion of the field, we identify a sharp peak near $x = 2d$, which can be attributed to the arrival of the third photon. This component exhibits bunching in both $g_a^{(2)}$ and $g_a^{(3)}$, as expected for a stimulated emission process involving three photons. The reduced width of the peak is consistent with an enhanced decay rate scaling as $\Gamma(N_{\rm lc} - 1)$, where $N_{\rm lc}$ denotes the number of emitters within the light cone at the given delay. 
  For this time delay, backscattering within the atomic array begins to generate feedback between emitter pairs, resulting in a noisier signal, as evident from the plotted photon correlations. This feedback marks the onset of in-phase, coherent light emission, accompanied by a gradual reduction of $g_a^{(2)}$ toward unity.

    \item Finally, for $x > 3d$, we observe the field reaching the detector at times $t > t_0+ 3\tau$. As in previous cases, the sharper peak can be attributed to the fast contribution arising from stimulated emission involving four photons. The remaining part of the signal can be interpreted as originating from in-phase emission resulting from feedback established first among triplets of emitters and eventually among all of the chain. Additionally, multiple backscattering events within the array  contribute to the emission visible in the tail. Indeed, the residual light escapes the edge of the chain predominantly in a bunched form, as single photons are reflected and experience the chain boundaries as effective mirrors.
    
\end{itemize}

It is worth noticing that this buildup of correlations stands in striking contrast to the Markovian case, represented by the gray line in Fig.~\ref{fig:figure_3}(c)-(d). In that scenario, an initial bunching ($x=0$ corresponds to $t=0$ in our setting) triggers the formation of a superradiant burst, as discussed in Ref.~\cite{PhysRevLett.131.033605}, but once established, the emitted field becomes predominantly coherent in the superradiant region. 
Finally, note that the bunching observed in the tail of the emission, in both the Markovian and non-Markovian cases, originates from relative fluctuations rather than from collective emission.

\subsubsection*{Entanglement generation} \label{sec:entanglement_generation}

We now examine how the two time-delayed effects introduced earlier, the modification of the superradiant burst and the excitation of BICs, impact the generation of entanglement in the system.
We first compute the entropy of entanglement between the emitters and the field
\begin{equation}
    S_{\mathrm{e-f}}(t) =-{\rm Tr}[\hat \rho_0(t) \log \hat\rho_0(t)] ,
\end{equation}
where $\hat \rho_0={\rm Tr}_{\rm f}[\hat \rho]$ is the reduced density matrix of the emitter system, obtained by tracing out the field degrees of freedom.
The time evolution of this entropy is shown in Fig.~\ref{fig:figure_4}(a). 
It exhibits a peak that coincides with the superradiant emission in the Markovian regime and, for large $\eta$, shifts toward the emitter half-life $t = -\log(0.5)/\gamma$, where the entanglement between a single atom and an emitted photon is maximized.
At long times, the atom-photon entanglement saturates to a finite value, corresponding to the entanglement between the atoms and the photons trapped into the BIC states, consistent with the discussion of the previous sections.

This saturation effect is also evident in the entanglement between the trapping region, defined as the subsystem consisting of the emitters and the segment of the waveguide between them, and the outer region, namely the rest of the waveguide. This quantity, $S_{\mathrm{in-out}}$, is plotted in Fig.~\ref{fig:figure_4}(b) and illustrates how this entanglement entropy increases with the time delay, as a larger portion of the photonic excitation, correlated with the emitted field, remains trapped between the emitters.
Note that this bipartition also captures the maximum amount of entanglement generated during the dynamics, which is relevant for choosing appropriate values of the maximum bond dimensions required by our MPS ansatz ($\chi \approx 2^{\max S_{\mathrm{in-out}}}$), as discussed, for the two-emitter case, in Ref.~\cite{PhysRevLett.116.093601}.

The BIC formation process, which emerges at long times, can also determine the entanglement generation between emitter pairs, which can be quantified by the half-chain logarithmic negativity~\cite{PhysRevLett.95.090503}, defined as
 $N_L=\log(\|\,\hat{\rho}_0^{T_L}\|_1)$ where $\hat{\rho}_0^{T_L}$ is the partial transpose of the emitters' reduced density matrix $\hat{\rho}_0$ with respect to the left half of the emitter chain, and $||\cdot||_1$ denotes the trace norm.
This quantity is a well-defined entanglement monotone that quantifies the bipartite entanglement between two subsystems of an open quantum system. Although it provides only a sufficient condition for entanglement, it offers a convenient  accessible alternative to more complete but less tractable measures, such as the entanglement of formation~\cite{PhysRevA.54.3824}.

We plot the logarithmic negativity in Fig.~\ref{fig:figure_4}(c), which reveals a striking behavior. In the Markovian regime, the logarithmic negativity remains identically zero, in agreement with recent findings that no entanglement is generated among emitters during superradiant decay~\cite{bassler2025absence,rosario2025unraveling,zhang2025unraveling}. In contrast, in the non-Markovian regime, a finite amount of entanglement does emerge, but crucially it appears only after the superradiant decay (indicated by the shaded gray region in the figure),  as the system relaxes into entangled BIC states.
Interestingly, the nature of this entanglement buildup depends on the propagation delay. For small $\eta$, entanglement grows smoothly. However, at larger delays, we observe distinct oscillations in the logarithmic negativity. On one hand, this behavior reflects the fact that the excited-state population of the emitters in a BIC tends to decrease with $\eta$, scaling like $\eta^{-1}$ at large values (see \cite{PhysRevA.94.043839,PhysRevLett.122.073601,magnifico2025non} for explicit expressions in the $N=2$ case). As a result, for large $\eta$ the photon excitation in the BIC becomes predominant,  with a reduced atomic component and therefore a smaller amount of atomic entanglement. 
On the other hand, in the large-delay regime the interaction between the emitted photon and a neighboring emitter occurs only after the emitting atom has almost completely decayed. This reduces the overlap between the emitted photon wave packet and the standing-wave profiles that characterize trapped photons in BIC states. Consequently, the photon undergoes multiple scattering events among the emitters (corresponding with the bounces observed in the figure), which lead to a gradual build-up of entanglement at intermediate delays, which gets 
progressively suppressed as $\eta$ increases, allowing excitations to eventually escape the array.
This behavior is also reflected in the pronounced excitation number oscillations shown in Fig.~\ref{fig:figure_2}.
Finally, we note that, as discussed in the next section, the generated entanglement originates from the trapping of excitations in the singlet subspace, which is spanned by an overcomplete basis of all possible dimer (singlet-state) coverings. This makes the entanglement among the emitters inherently multipartite and dimerized.

\subsection*{Excitation trapping}\label{SecBIC}

\begin{figure}
    \centering
    \includegraphics[width=0.99\linewidth]{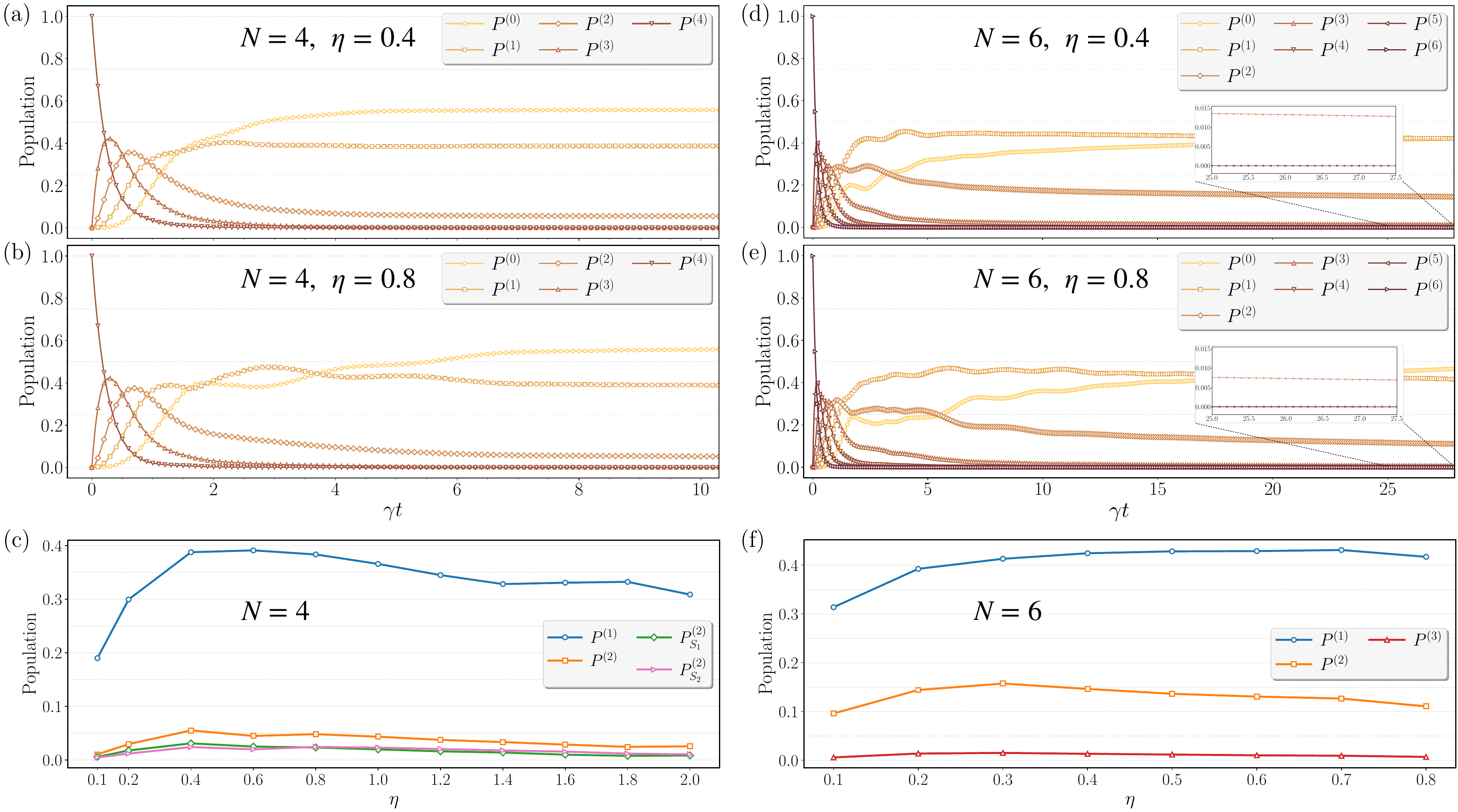}
    \caption{Populations $P^{(N_{\mathrm{exc}})}$ of the emitters' Hilbert space sectors characterized by different excitation numbers at $\eta=0.4$ and  $\eta=0.8$ for $N=4$ qubits [panel (a) and panel (b)] and $N=6$ qubits [panel (d) and panel(e)]. In panel (c), for $N=4$, we report as a function of $\eta$ the populations of the $N_{\mathrm{exc}}=1$ and $N_{\mathrm{exc}}=2$ excitations subspaces, and of the two states in Eqs.~\eqref{eq:singl1}-\eqref{eq:singl2}, which span the singlet subspace. The population values are computed in the long-time limit ($\gamma t=16$). In panel (f), we report the populations of the $N_{\mathrm{exc}}=1$, $N_{\mathrm{exc}}=2$ and $N_{\mathrm{exc}}=3$
    excitations subspaces for the case with $N=6$. The population values are computed in the long-time limit ($\gamma t=25$). }
    \label{fig:BICS_analyses}
\end{figure}

We now analyze in detail the trapping of part of the initial excitation in the long-time relaxation dynamics, already described in ``\nameref{Sec.noMark}''.
In the $N_{\mathrm{exc}}=1$ subspace, this effect reflects the formation of BIC states predicted for any number of equally spaced two-level emitters~\cite{Lonigro_2021, PhysRevLett.122.073601, papaefstathiou2024efficient,PhysRevResearch.6.023213}. 
For sectors with $N_{\mathrm{exc}}>1$, recent works have addressed the presence of subradiant dimerized states at half-filling~\cite{PhysRevLett.127.173601,PhysRevA.110.053707,1r6c-6ftv}. However, the exploration of higher-excitation subspaces via relaxation processes is still open.

To investigate whether the excitation trapping observed in our simulations also involves higher-excitation subspaces, we show in Fig.~\ref{fig:BICS_analyses} the emitter populations in the different $N_{\mathrm{exc}}$ sectors as a function of time, for two values of $\eta$ in chains of $N=4$ and $N=6$ emitters. We find that no population is retained in the $N_{\mathrm{exc}} > N/2$ subspaces, which instead decay subexponentially, while asymptotic trapping is clearly observed in the $N_{\mathrm{exc}} \leq N/2$ excitation sectors. Interestingly, the trapping into the two excitation sector results enhanced for the larger array with $N=6$ atoms given the increased dimension of the corresponding dark singlet subspace. 

Half-filling marking a threshold for the maximum number of trapped excitations is an expected feature for all even $N$. In the Markovian regime described by the master equation~\eqref{eq:ME_Dicke}, this property can be understood by the fact that states protected from decay (namely, annihilated by the total ladder operator $\hat{S}$) are the ones fulfilling $S_z=-S$, corresponding to
\begin{equation}
    N_{\mathrm{exc}} = \frac{N}{2} + S_z = \frac{N}{2} - S \in \left\{ 0,1,\dots , \frac{N}{2} \right\} .
\end{equation}
The case $S=N/2$  corresponds to the trivial ground state ($N_{\mathrm{exc}}=0$), while the states with $N_{\mathrm{exc}}>0$ are closely connected to BICs in the Hamiltonian picture. Stable half-filled states exist for arbitrary even $N$, as they coincide with the singlet subspace ($S = S_z = 0$) of dimension  $N!/[(N/2)!(N/2+1)!]$ as obtained from the Clebsch-Gordan decomposition. This subspace can be spanned by an overcomplete basis consisting of all possible dimer (singlet-state) coverings.
However, these states are inaccessible to a Markovian dynamics that starts from the initial state~\eqref{eq:Nexc}, which remains confined in the $S=N/2$ total spin manifold, and can be populated only due to non-Markovian effects.

In Fig.~\ref{fig:BICS_analyses}, we show how the trapping in the first two excitation subspaces depends on the time delay, with the average maximum trapping observed at $\eta \sim 0.4$. Since this trapping is at the origin of the generated entanglement, such a delay also leads to a maximum in the entanglement in the long-time limit.
To investigate which states are involved in this trapping, we consider the following two states spanning the two-fold degenerate singlet subspace with $S_z=S=0$:
\begin{equation}\label{eq:singl1}
    \ket{\psi^{(2)}_{S_{1}}}=\frac{\ket{e,e,g,g}-\ket{e,g,e,g}-\ket{g,e,g,e}+\ket{g,g,e,e}}{2},
\end{equation}
and
\begin{equation}\label{eq:singl2}
    \ket{\psi^{(2)}_{S_{2}}}= \frac{\ket{e,e,g,g}+\ket{e,g,e,g}+\ket{g,e,g,e}+\ket{g,g,e,e}}{2\sqrt{3}} -\frac{\ket{e,g,g,e}+\ket{g,e,e,g}}{\sqrt{3}}.
\end{equation}
Their populations in the long-time limit 
\begin{equation}
    P^{(2)}_{S_{\ell}}=\bra{ \psi^{(2)}_{S_{\ell}}} \hat{\rho}_{0}(t\to \infty) \ket{\psi^{(2)}_{S_{\ell}}} \quad \text{with } \ell=1,2
\end{equation}
are reported in Fig.~\ref{fig:BICS_analyses}(c). We observe that the sum of the two projections equals the population of the two-excitation subspace $P^{(2)}$, indicating that the excitations retained in this sector originate from trapping into the singlet subspace. 

\begin{figure}
    \centering
    \includegraphics[width=0.7\textwidth]{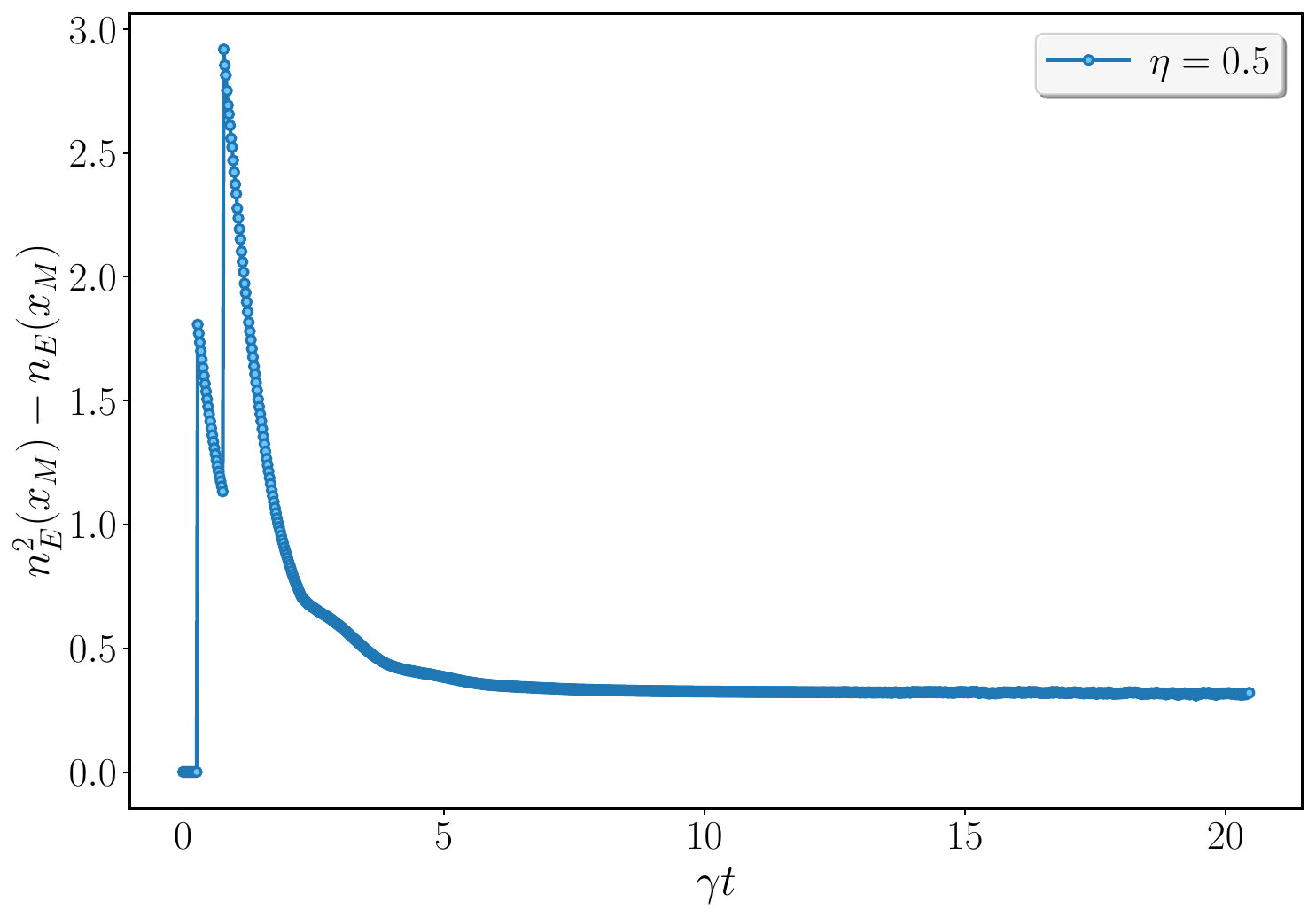}
    \caption{Time evolution of the quantity $n_E^2(x_M) - n_E(x_M)$, where $x_M$ is the midpoint of the intra-emitter region, computed for the case with $N=4$ emitters. The initial state of the system is reported in Eq.~\eqref{eq:Nexc} while the delay corresponds to $\eta=0.5$.}
    \label{fig:occupation_num_N_4_eta_0.5}
\end{figure}

We note that the non-Markovian decay for the limiting case of two initially excited emitters was analytically studied in Ref.~\cite{PhysRevResearch.6.023213}. In addition to the trapping in the single-excitation BIC, which we also observe, the authors reported the emergence of a finite two-photon component trapped by the atoms in the long-time limit.
To investigate if this also occurs in our case where more emitters and photons are involved, we computed the quantity $n_{E}^2(x_M) - n_{E}(x_M)$, where $x_M$ denotes the midpoint of the array, for $N=4$ emitters. As reported in Fig.~\ref{fig:occupation_num_N_4_eta_0.5}, we found that this quantity approaches a finite asymptotic value around $0.32$, thereby confirming the presence of multiphoton trapping between the two central emitters as observed for two atoms in Ref.~\cite{PhysRevResearch.6.023213}. A detailed analysis of the nature, origin, and dependence of this effect on the number of excitations and emitters is left for future work.

\subsection*{Enhanced superradiant decay}\label{Sec.Supersuper}

\begin{figure}
    \centering
    \includegraphics[width=0.92\textwidth]{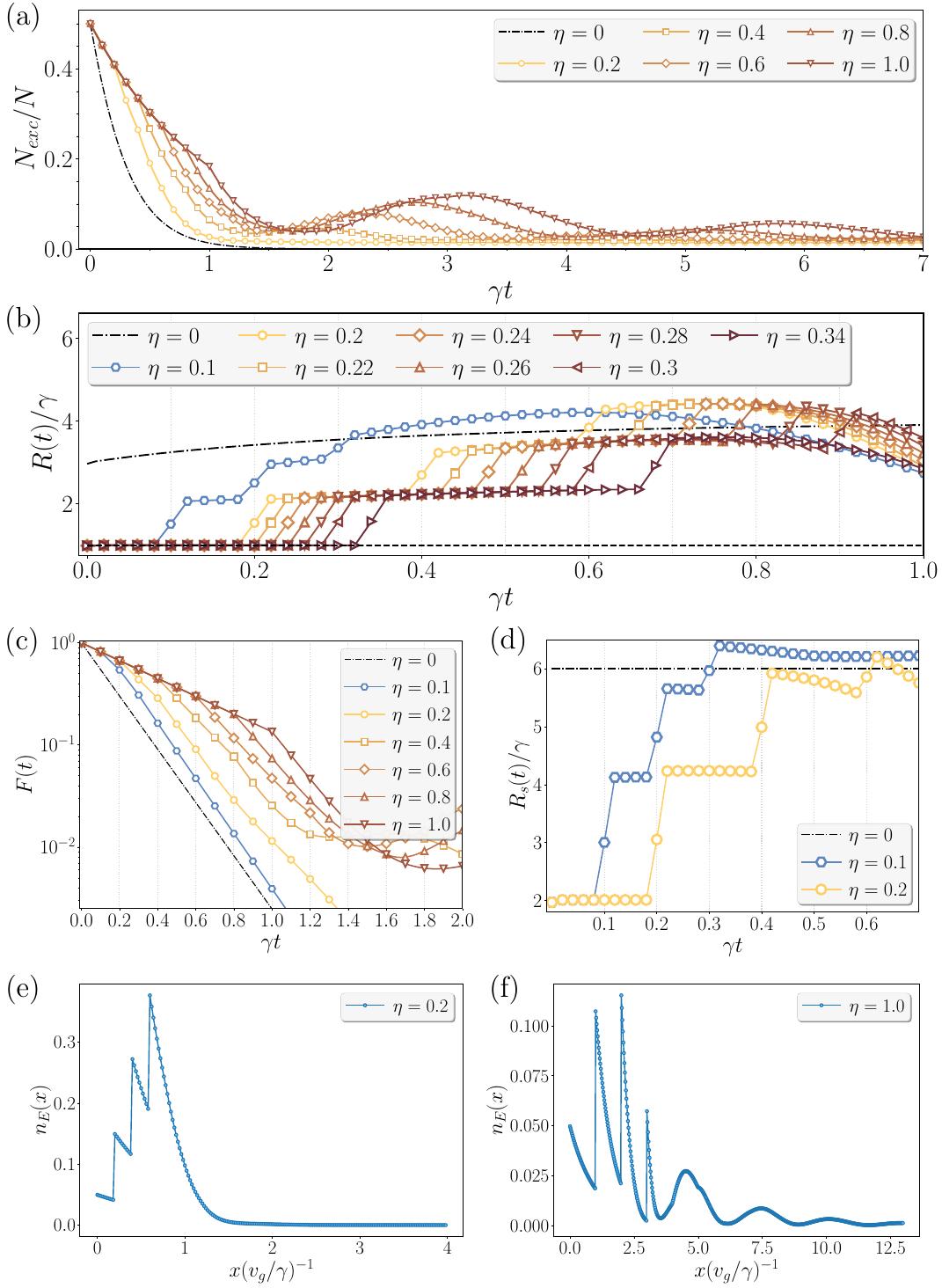}
    \caption{Decay dynamics of a totally symmetric Dicke state with $N_{\mathrm{exc}}=N/2$. \textit{Panel (a).} Time evolution of the emitter excitation number for an array with $N = 4$, prepared in the symmetric state~\eqref{eq_sym}, for different values of $\eta$.
    \textit{Panel (b).} Plot of the instantaneous decay rate extracted from data, compared with the Markovian case at $\eta = 0$ (dot-dashed line), obtained by solving the master equation~\eqref{eq:ME_Dicke}, and with the result $R(t)/\gamma=1$ for independent emitters (dashed horizontal line).
    \textit{Panel (c).} Time evolution of the survival probability of the symmetric state~\eqref{eq_sym}. \textit{Panel (d).} Instantaneous decay rates for the initial state~\eqref{eq_sym}; the dashed horizontal line indicates the Markovian limit result $R(t)=6\gamma$. \textit{Panels (e)-(f).} Long-time snapshot of the field energy density profile along the waveguide, for two different values of $\eta$.}
    \label{fig:fig5}
\end{figure}

We complete our analysis by considering an initial configuration that differs from the one discussed previously, where all atoms were initially excited. Specifically, we consider the state
\begin{equation}\label{eq_sym}
    \ket{\Psi(0)}_s=|S=N/2,S_z=0\rangle\otimes|{\rm vac}\rangle
\end{equation}
where the field is in the vacuum state and the emitters are prepared in the totally symmetric Dicke state with $N_{\mathrm{exc}}=N/2$. In the Markovian regime, this state has the highest decay rate in the system, given by \mbox{$\Gamma_s= N/2(N/2+1)\gamma$~\cite{GROSS1982301}}.
For $N=2$, it was shown that, at certain values of the propagation delay, this state can decay at a rate that exceeds the expected value of \mbox{$\Gamma_s=2\gamma$}, an effect dubbed as ``superduperradiance''~\cite{PhysRevLett.124.043603,PhysRevResearch.1.032042,PhysRevA.102.043718}.
To investigate whether this effect persists at higher excitation fillings, in Fig.~\ref{fig:fig5}(a) we consider the decay of the symmetric state~\eqref{eq_sym} and we plot the average excitation population of the emitters defined in Eq.~\eqref{eq_pmean} for various time delays.
We observe that for $\eta>0$, the system initially undergoes a slower decay, given by independent emission. However, once more emitters enter the same light cone, collective in-phase emission begins to build up, leading to a transient regime in which the decay rate exceeds that predicted in the Markovian limit. For the case of $N=4$ emitters, this occurs for $t > 3\tau$ (i.e.\ $\gamma t > 3 \eta$), when sufficient feedback has established in-phase emission. This effect is clearly visible in Fig.~\ref{fig:fig5}(b), where we plot the instantaneous decay rate, defined in Eq.~\eqref{eq:rate}, as a function of time for different values of $\eta$. We observe that the maximum enhancement in the decay is obtained for $\eta\sim 0.23$. Note that in Ref.~\cite{PhysRevLett.124.043603}, which considers $N=2$ emitters, the optimal delay was estimated to be $\eta \sim 0.56$. This suggests that the optimal delay may decrease with the number of atoms. However, a systematic investigation of its dependence on the number of emitters is required to clearly assess this behavior.

To definitely demonstrate that the ``superduperradiance'' effect indeed arises from time-delayed feedback, which allows the symmetric state~\eqref{eq_sym} to decay at a rate faster than that predicted by the Markov approximation ($\Gamma_s = 6\gamma$), we plot in Fig.~\ref{fig:fig5}(c) the survival probability of the symmetric state, $F(t) = |_s\langle \Psi(0) | \Psi(t) \rangle_s|^2$. From this quantity, we extract the instantaneous decay rate, $R_s(t) = -\dot{F}(t)/F(t)$, plotted as a function of time in the right panel of the figure. The plot clearly shows that the delayed feedback can indeed produce a decay faster than the Markovian prediction (dashed line), thereby extending the results of Refs.~\cite{PhysRevLett.124.043603,PhysRevResearch.1.032042,PhysRevA.102.043718} to higher excitation sectors.

Finally, we also consider the emitted field, plotted in Fig.~\ref{fig:fig5}(d), where we observe the appearance of multiple energy density peaks, as shown previously in Fig.~\ref{fig:figure_2}(c)--(d). Compared to the case where all emitters are initially excited, the symmetric initial state considered here leads to a stronger build-up of the emitted light, although without exceeding the peak intensity of the Markovian superradiant burst.

\section*{Discussion}

In this work, we investigated the fate of superradiant decay in the presence of non-negligible propagation delays. Using a tensor-network-based method, we simulated the many-body dynamics of atomic ensembles with up to eight emitters, explicitly accounting for non-Markovian effects due to delayed photon exchange. Our results reveal that these effects can play a constructive role, shifting the conventional picture of superradiance from coherent in-phase emission to the generation of highly correlated photonic states. These correlations, as previously observed in the Markovian regime~\cite{PhysRevLett.116.093601,PhysRevA.93.062104,PhysRevX.14.031043}, do not seem to generate entanglement among the emitters during superradiant decay, while entanglement emerges at later times, as the system relaxes towards bound states in the continuum with $N_{\mathrm{exc}}\leq N/2$. Finally, we demonstrated that for the most radiant configuration, the symmetric Dicke state, the decay dynamics can transiently exceed the Markovian decay rate, extending previous results~\cite{PhysRevLett.124.043603,PhysRevResearch.1.032042,PhysRevA.102.043718} to higher emitter and exciation numbers.

The phenomenology described in this article is already within reach of state-of-the-art non-Markovian waveguide QED experiments, such as those based on circuit QED and matter-wave platforms. In particular, circuit QED platforms have demonstrated the ability to reach large time delays \mbox{(\(\eta \gg 1\))~\cite{PhysRevX.11.041043,ferreira2024deterministic},} with negligible dissipation into external decay channels, remaining below 1\% compared to emission into the waveguide~\cite{PhysRevX.11.041043,ferreira2024deterministic,PhysRevX.12.031036} and support for coupling multiple emitters ($N > 2$), which is essential for observing the superradiant burst~\cite{zhang2023superconducting}. 
Emitters interacting via matter waves also represent a promising platform to test our predictions. In this setting, sub- and superradiant interactions at small time delays ($\eta \approx 0.1$) were demonstrated for multiple emitters, revealing the onset of delayed collective dynamics and a small population trapping, attributable to the formation of a bound state in the continuum~\cite{kim2025super}. We emphasize that our predictions, particularly the formation of a correlated train of photons, are actually enhanced at intermediate delays ($\eta \lesssim 1$) and remain clearly observable even at the small delays achieved in current experiments.
In platforms that suffer from significant dissipation into external modes, such as cold-atom and solid-state emitter systems, the trapping of excitations and the resulting entanglement generation would be strongly affected, as they would no longer be protected by a perfect dark state. On the other hand, external losses could lead to qualitatively different phenomenology, particularly during the correlated emission process.
Efficiently incorporating external dissipation within our numerical framework remains an open challenge, which we plan to address in future work.

During preparation of this work, we became aware of two related studies: one by C.~Barahona-Pascual et al.~\cite{barahona2025time}
and another by S.~Arranz Regidor et al.~\cite{regidor2025theory}. The first study reported a fragmentation of the photon burst in the output field, consistent with our observations. The second focused on photon-photon and atom-photon correlations in a system with two and four emitters, focusing on two-excitation states and also discussing the trapping into multi-excitation BICs. The results in these articles integrate and corroborate our main findings, namely the emergence of a correlated photon train, the development of entanglement among emitters during time-delayed superradiant decay, the trapping of multiple excitations into the singlet entangled subspace, and the enhanced superradiant decay of the symmetric Dicke state.

We finally remark that the numerical framework developed in this work, along with our results, opens several intriguing directions for future research. In particular, the method could be extended to study the scattering of incoming photon pulses propagating through the waveguide  with the emitter array, as well as to investigate the steady-state  of the system under continuous external driving.
The latter scenario would allow for the exploration of the driven Dicke model in the presence of time delay,  a scenario  that has recently attracted considerable interest in the Markovian regime~\cite{ferioli2023non,PRXQuantum.5.040335,PRXQuantum.6.020303,Glicenstein:22,shimshi2024dissipative}.

\section*{Methods}\label{Appendix:A}

\subsection*{MPS encoding and evolution}
\label{Appendix:MPS_simulations}

To encode and evolve the quantum state of the qubit-field system as a matrix product state (MPS), we generalize the methodology and algorithms presented in~\cite{PhysRevLett.116.093601} to accommodate systems with many emitters. 

The interaction Hamiltonian of $N$ identical quantum emitters, equally spaced by a distance $d$, in a one-dimensional waveguide can be written in the Schr\"odinger picture as in Eq.~\eqref{eq:V_continuousTime}. To build the MPS representation, it is convenient to shift the left-propagating quantum noise operators~\eqref{eq:noise_operators} as follows:
\begin{equation}
    \hat b_L(t) \rightarrow e^{i(N-1)\phi} \hat b_L\left(t-(N-1)\tau\right),
\end{equation}
so that Eq.~\eqref{eq:V_continuousTime} becomes
\begin{equation} \label{eq:shifted_Ham}
    \tilde{H}^{\mathrm{int}}_{I}(t) = \hbar \sqrt{\frac{\gamma}{2}} \sum_{j=1}^N
    \hat{\sigma}_j\Bigl[ e^{-i(j-1)\phi}\hat b_R^\dag(t-(j-1)\tau) + e^{-i(N-j)\phi}\hat b_L^\dag(t-(N-j)\tau)\Bigr]  + \mathrm{H.c.}\, , 
\end{equation}
where $\tau=d/v_g$ is the time delay in the photon propagation from the $k$-th emitter to the $(k+1)$-th, as discussed in ``\nameref{Sec.model}'', and $\phi=\omega_0\tau=k_0 d$. After coarse-graining the evolution (see ``\nameref{Sec.model}''), the interaction Hamiltonian~\eqref{eq:shifted_Ham} leads to the dynamical map
\begin{equation} \label{eq:dynamical_map_MPS}
   \hat U_n=\prod_{j=1}^{N/2} e^{-i\hat O_{n,j}}, 
\end{equation}
with
\begin{align}\label{eq:3legs_op_MPS}
    \hat O_{n,j} & =\sqrt{\frac{\gamma \Delta t}{2}}\left\{\hat{\sigma}_j \left[ e^{-i(j-1)\phi}\hat B^\dagger_{n-(j-1)\ell,R} + e^{-i(N-j)\phi}\hat B^\dagger_{n-(N-j)\ell,L}\right] \right. \\ 
    & 
    + \left. \hat{\sigma}_{N-(j-1)} \left[ e^{-i(N-
    j)\phi}\hat B^\dagger_{n-(N-j)\ell,R} + e^{-i(j-1)\phi}\hat B^\dagger_{n-(j-1)\ell,L} \right] \right\} + \mathrm{H.c.},
\end{align} 
where the coarse-graining interval $\Delta t$ is conveniently chosen to be a unit fraction of the delay, in such a way that $\tau=\ell\Delta t$ and $\ell\in\mathbb{N}$.\\
Therefore, at each time step of the evolution, $t_n\rightarrow t_{n+1}=t_n+\Delta t$, the state of the full system evolves according to
\begin{equation} \label{eq:collision_Un_MPS}
    \ket{\Psi(t_{n+1})} = \hat U_n\ket{\Psi(t_{n})} = \exp\Bigl \lbrace-i  \sum_{j=1}^{N/2} \hat O_{n,j}  \Bigr\rbrace \ket{\Psi(t_{n})},
\end{equation}
which has the same structure of  Eq.~\eqref{eq:collision_Un}.\par
The MPS scheme for the non-Markovian collisional evolution is based on the iterated application of the dynamical map~\eqref{eq:collision_Un_MPS}, after encoding the initial state of the system into an MPS whose sites represent both emitters and photonic degrees of freedom.

\begin{figure}
    \centering
    \includegraphics[width=0.90\textwidth]{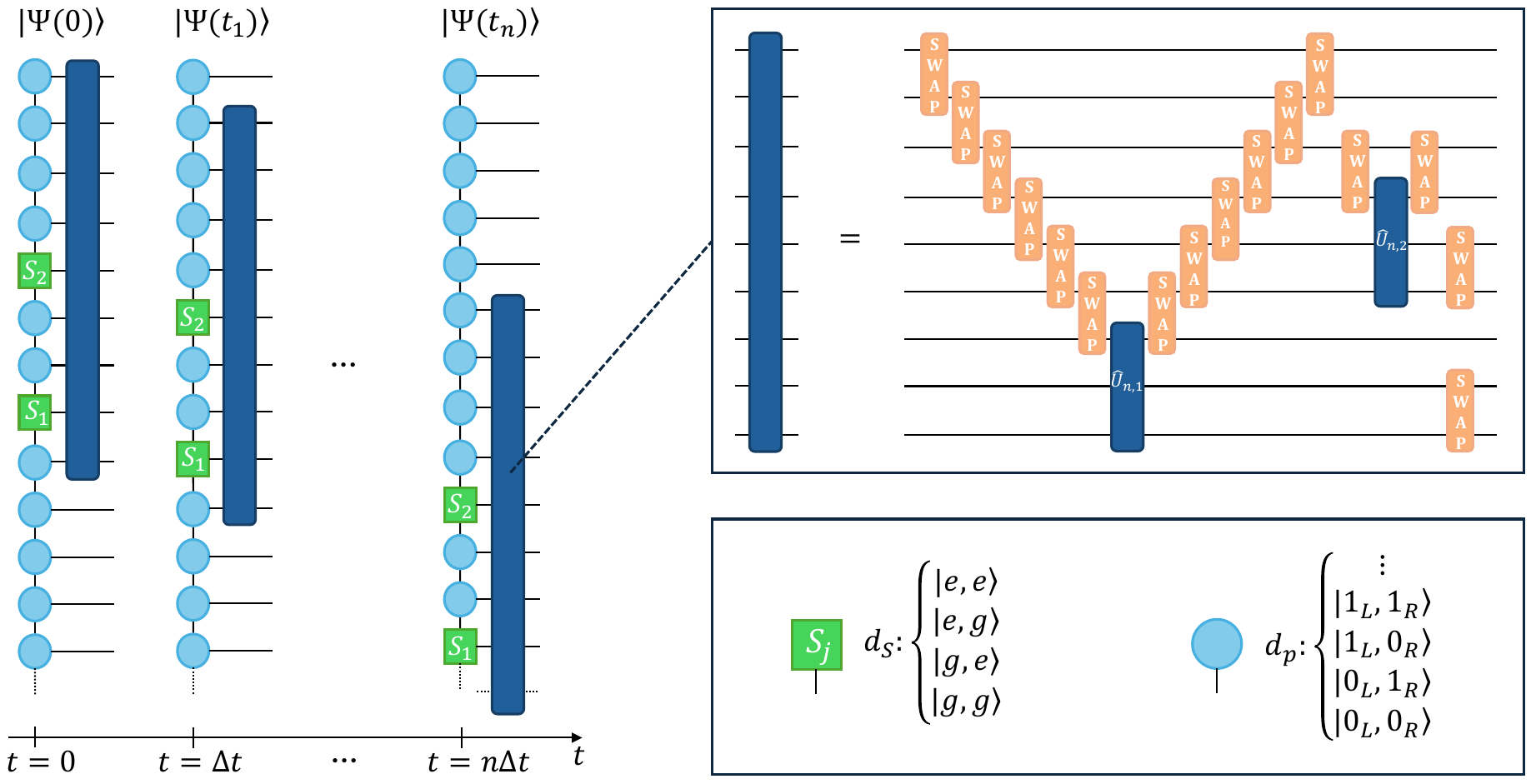}
    \caption{Tensor Network simulation scheme for the non-Markovian dynamics of the waveguide QED system, shown for $N = 4$ emitters and $\ell=2$. The corresponding value of the dimensionless parameter is $\eta=\gamma \tau = \gamma \ell \Delta t$. The emitters are grouped in pairs at sites $S_1$ and $S_2$ (green), separated by photonic sites (light blue) encoding the collision units. The evolution from $\ket{\Psi(0)}$ to $\ket{\Psi(t_{n})}$ is performed through iterative application of unitary gates $\hat{U}_{n,j}$ (blue) combined with SWAP operations (orange) to handle non-local interactions due to the time delays.} 
    \label{fig:figure_MPS}
\end{figure}

To this end, let us note that Eqs.~\eqref{eq:dynamical_map_MPS}--\eqref{eq:collision_Un_MPS} provide a key to arrange the emitters in the MPS representation. In fact, at each time step of the evolution, each quantum emitter interacts \emph{only} with two different collision units. In particular, at the $n$-th time step, emitter $j$ and emitter $N-(j-1)$ interact with the collision units encoded in the  time-bins $n-(j-1)\ell$ and $n-(N-j)\ell$. This holds for $j=1,\ldots,N/2$, assuming that the $N$ (even) emitters are ordered from left to right. Hence the dynamics of the emitters can be decomposed into independent pairs $\{j, N-(j-1)\}$, since the unitary operators $\hat U_{n,j}$~\eqref{eq:dynamical_map_MPS}, which act on each emitter pair at every time step, mutually commute. 
Thus, when constructing the MPS, we introduce $N/2$ emitter sites, $S_1,\ldots,S_{N/2}$, where each site $S_j$ hosts the emitter pair $\{j, N-(j-1)\}$. Moreover, in order to encode the photonic degrees of freedom in a non-Markovian situation, we have to consider a finite time delay $\tau$ in the photon propagation between two consecutive emitters. In the MPS representation, this is translated into separating the emitter sites by $\ell=\tau/\Delta t$ photonic sites, each encoding both left- and right-propagating collisional photonic modes. However, since there are $N$ emitters in the system, but only $N/2$ emitter sites in the MPS scheme, it is necessary to take into account the remaining $N/2$ intra-emitter photonic degrees of freedom. These are encoded into $N \ell /2$ photonic sites on the right-hand side of the final emitter node $S_{N/2}$. In addition, the collision units representing  left- and right-propagating photons in the extra-emitter region are encoded in the photonic sites on the left side of the first emitter node $S_1$. For a site $S_j$ containing a pair of quantum emitters, the Hilbert space is $\mathcal{H}_{S_j}$ with dimension $d_S$, whereas $\mathcal{H}_{p}$ is the Hilbert space of the photonic sites, hosting the collisional field modes, with dimension $d_p$.

After building the initial MPS as discussed above, we simulate the dynamics~\eqref{eq:collision_Un_MPS} by exploiting the commutation property of the unitary operators $\hat U_{n,j}$ to decompose the total evolution operator $\hat U_n$ as in Eq.~\eqref{eq:dynamical_map_MPS}. We apply the operators $\hat U_{n,j}$ orderly, with $j=1,\ldots,N/2$, noticing that each $\hat U_{n,j}$ acts on one emitter site, $S_j$, and two photonic time-bins, $n-(j-1)\ell$ and $n-(N-j)\ell$. Thus, evolution features long-range interactions, which can be modeled by performing a series of $\mathrm{SWAP}$ operations to move the MPS photonic sites, involved in the interaction, next to the emitter site $S_j$. In this way, the unitary $\hat U_{n,j}$ can be applied locally to the three sites in question, that are now contiguous. After applying $\hat U_{n,j}$, we restore the MPS form of the evolving state by performing a singular value decomposition (SVD) and we swap back the photonic bins to their original position. We repeat this approach for each $j=1,\ldots,N/2$. Afterwards, we prepare the state for the next time step by shifting each emitter site of one position in the MPS chain by means of $\mathrm{SWAP}$ gates and we iterate the procedure described above to evolve the state according to Eq.~\eqref{eq:collision_Un_MPS}. This simulation technique is depicted in Fig.~\ref{fig:figure_MPS}, for the case of $N=4$ emitters and  $\ell=2$.

In the study of emitter collective effects in the non-Markovian regime (Figs.~\ref{fig:figure_2},~\ref{fig:fig5}), and of the behavior of entanglement  (Fig.~\ref{fig:figure_4}), we performed MPS simulations using $d_S=d_p=4$ as physical dimensions of emitter and photonic sites. Each site's four-dimensional Hilbert space decomposes as $\mathcal{H}_S=\mathcal{H}_p\simeq\mathbb{C}^2\otimes\mathbb{C}^2$: for emitter sites, this encodes two qubits; for photonic sites, this encodes left- and right-propagating modes, each taking values 0 or 1. We note that it is possible to neglect the higher occupation levels of the bosonic modes, since the probability of having $m$ excitations in each photonic bin scales as $(\Delta t)^m$~\cite{PhysRevX.14.031043}. Hence, for sufficiently small $\Delta t$, the leading contributions come from $m=0,1$.

However, for studying photon auto-correlation functions of order $m$, as described in ``\nameref{sec:photon_corr}'', it is necessary to take into account higher occupation levels of the photonic modes. In particular, to compute $G_a^{(2)}$ and $G_a^{(3)}$, defined in Eq.~\eqref{eq_correlation} and shown in Fig.~\ref{fig:figure_3}, we had to take into account higher-order contributions for $m=2,3$. In this case, $\mathcal{H}_p\simeq\mathbb{C}^4\otimes\mathbb{C}^4$. Therefore, we performed the simulations by increasing the physical dimension of the photonic sites in the MPS to $d_p=16$. To maintain uniform MPS structure, we also expanded the physical dimension of the emitter sites to $d_S=16$, though only the two-qubit subspace is physically relevant.

Numerical simulations were performed on the ReCaS computing cluster~\cite{ReCaS}.

\subsection*{Initial state and density matrix}
\label{Appendix:Initial_State}
In the MPS simulations, we considered two different initial states, depending on the analysis being performed.

In ``\nameref{Sec.noMark}'' and ``\nameref{SecBIC}'', we considered the situation where at $t=0$ the waveguide is empty and all emitters are excited. In this case, the initial state of the full system is completely separable, allowing us to directly build its MPS representation with initial bond dimension $\chi_{\rm t=0}=1$, encoding in each site either two excited emitters or empty photonic modes.

In ``\nameref{Sec.Supersuper}'', the initial state of the full system is defined in Eq.~\eqref{eq_sym}: the field is in the vacuum and the $N=4$ emitters are prepared in the totally symmetric Dicke state with half of the emitters excited and half in the ground state, i.e.,
\begin{equation} \label{eq:entangled_state}
    \ket{\psi_{D}}=\frac{1}{\sqrt{6}} \Bigl(\ket{e,e,g,g}+\ket{e,g,e,g} +\ket{e,g,g,e} + \ket{g,e,e,g}+\ket{g,e,g,e}+\ket{g,g,e,e}\Bigr).
\end{equation}
This state is entangled, due to the initial quantum correlations of the emitters. To prepare this state in the MPS, we initially create a fully separable state as discussed earlier, encoding the vacuum in each photonic site and temporarily filling the emitter nodes locally with randomly chosen configurations. Using $\mathrm{SWAP}$ gates, we bring the emitter sites close to each other. Starting from the state vector~\eqref{eq:entangled_state}, we reshape it to match the structure with two emitters per site and decompose it using exact SVD to obtain two tensors connected by an internal bond. These tensors are then plugged into the MPS, replacing the initial random tensor of the emitter sites. Finally, we bring back the emitter nodes to their original positions in the MPS chain.

A similar procedure can be used to compute the density matrix of the system composed of all the emitters at each time step of the simulation, as done in ``\nameref{sec:entanglement_generation}'' and ``\nameref{SecBIC}''. For this purpose, we bring the emitter nodes next to each other through $\mathrm{SWAP}$ operations and, after isolating them from the rest of the tensor network, we properly contract the so-obtained reduced MPS with its adjoint. The resulting tensor is then reshaped into a matrix, corresponding to the density operator of the emitter system.

\begin{figure}
    \centering
    \includegraphics[width=\textwidth]{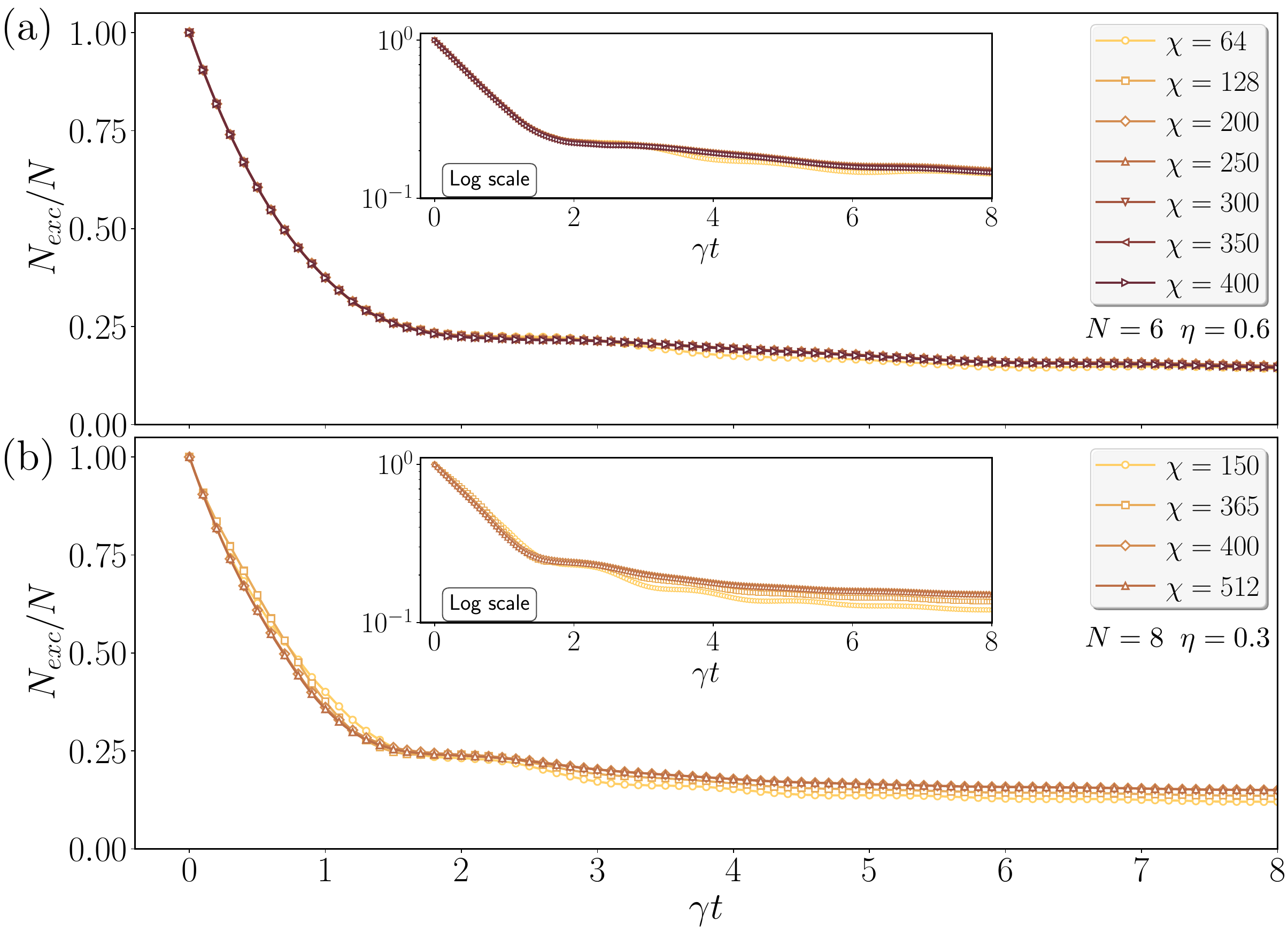}
    \caption{Test of numerical convergence on the behavior of the mean excitation population when varying the maximum bond dimension value $\chi$ in the simulation, at fixed number of emitters $N$ and delay parameter $\eta$; respectively (a) $N=6$, $\eta=0.6$, and (b) $N=8$, $\eta=0.3$.} 
    \label{fig:fig_convergence}
\end{figure}

\subsection*{Numerical convergence}
\label{Appendix:Num_convergence}

For the numerical simulations, we used bond dimensions that ensure to accurately capture the entanglement scaling examined in ``\nameref{sec:entanglement_generation}'', as a function of the number of emitters $N$ and the delay parameter $\eta=\gamma\tau$, defined in ``\nameref{Sec.model}''. Across the various system configurations studied, we employed bond dimensions up to $\chi=512$ for the most demanding cases. To ensure stability of our findings, we verified the convergence of the computed observables, e.g., the average excitation population $N_{exc}(t)$~\eqref{eq_pmean}, as a function of the bond dimension $\chi$, as shown in Fig.~\ref{fig:fig_convergence}.

Furthermore, depending on the specific simulations performed, we set the dimensionless coarse-graining time interval $\gamma \Delta t$ to values in the range $0.002$--$0.02$.

\backmatter





\bmhead{Acknowledgements}

The authors acknowledge computational resources provided by the University of Bari-INFN cluster ReCaS (\url{https://www.recas-bari.it/}).

\bmhead{Funding}

The authors acknowledge support from INFN through the project ``QUANTUM" and from the Italian funding within the ``Budget MUR - Dipartimenti di Eccellenza 2023–2027" - Quantum Sensing and Modelling for One-Health (QuaSiModO).
PF acknowledges support from the Italian National Group of Mathematical Physics (GNFM-INdAM) and from PNRR MUR project CN00000013-``Italian National Centre on HPC, Big Data and Quantum Computing``. GM acknowledges support from the University of Bari via the 2023-UNBACLE-
0244025 grant and from INFN through the project ``NPQCD". MM, SP and FVP acknowledge support from PNRR MUR project PE0000023-NQSTI. FVP acknowledges support from the project PRIN 2022 ``Emerging gauge theories: critical properties and quantum dynamics'' (20227JZKWP). GC and SM acknowledge that results incorporated in this standard have received funding from the T-NiSQ consortium agreement financed by QUANTERA 2021, via the Quantum Technology Flagship projects
PASQuanS2.1, the WCRI Quantum Computing and Simulation
Center (QCSC) of Padova University and  by the Italian  Ministry of University and Research MUR Departments of Excellence grant 2023-2027 ``Quantum Frontiers'' (FQ). 

\bmhead{Competing interests}

The authors declare no competing interest.

\bmhead{Data availability}

The data that support the findings of this study, including those used to generate Figs.~\ref{fig:figure_2}, \ref{fig:figure_3}, \ref{fig:figure_4}, \ref{fig:BICS_analyses}, \ref{fig:occupation_num_N_4_eta_0.5}, \ref{fig:fig5}, \ref{fig:fig_convergence}, are available from the corresponding author upon request.

\bmhead{Author contribution}

R.L.C., G.M. and M.M. developed the algorithm. R.L.C. performed numerical simulations and data analysis. G.C., F.V.P, M.M. and G.M. designed the study. S.M., S.P. and P.F. supervised the project. All authors discussed and validated the results, contributed to the writing of the manuscript, and approved its final version.



\end{document}